\documentclass[11pt,twocolumn]{IEEEtran_v15}
\usepackage[dvips]{graphicx}
\usepackage{myusenix}

\usepackage{amsmath}
\usepackage{url}

\begin{document}

\renewcommand\dbltopfraction{.95}
\renewcommand\textfraction{.05}

\newcommand{\eat}[1]{} 

\title{Practical Load Balancing for Content Requests in Peer-to-Peer Networks}

\author{
Mema Roussopoulos \hspace{0.75cm} Mary Baker\\ 
{\em Department of Computer Science} \\
{\em Stanford University}\\
{\em Stanford, California, 94305}\\
\\
\normalsize \{mema, mgbaker\}@cs.stanford.edu \\
 http://mosquitonet.stanford.edu/} 

\maketitle 
\begin{abstract}
This paper studies the problem of load-balancing the demand for
content in a peer-to-peer network across heterogeneous peer nodes that
hold replicas of the content.  Previous decentralized load balancing
techniques in distributed systems base their decisions on periodic
updates containing information about load or available capacity
observed at the serving entities.  We show that these techniques do
not work well in the peer-to-peer context; either they do not address
peer node heterogeneity, or they suffer from significant load
oscillations.  We propose a new decentralized algorithm, Max-Cap,
based on the maximum inherent capacities of the replica nodes and show
that unlike previous algorithms, it is not tied to the timeliness or
frequency of updates. Yet, Max-Cap can handle the heterogeneity of a
peer-to-peer environment without suffering from load oscillations.

\end{abstract}

\section{Introduction}

Peer-to-peer networks are becoming a popular architecture for content
distribution \cite{oram01}.  The basic premise in such networks is that any
one of a set of ``replica'' nodes can provide the requested content,
increasing the availability of interesting content without requiring
the presence of any particular serving node.

Many peer-to-peer networks push index entries throughout the overlay
peer network in response to lookup queries for specific
content \cite{gnutella,ratnasamy01a,rowstron01b,stoica01,zhao01}.
These index entries point to the locations of replica nodes where the
particular content can be served, and are typically cached for a finite
amount of time, after which they are considered stale.  Until now,
however, there has been little focus on how an individual peer node
should choose among the returned index entries to forward client
requests.

One reason for considering this choice is load balancing. Some replica
nodes may have more capacity to answer queries for content than
others, and the system can serve content in a more timely manner by
directing queries to more capable replica nodes.

In this paper we explore the problem of load-balancing the demand for
content in a peer-to-peer network.  This problem is challenging for
several reasons.  First, in the peer-to-peer case there is no
centralized dispatcher that performs the load-balancing of requests;
each peer node individually makes its own decision on how to allocate
incoming requests to replicas.  Second, nodes do not typically know
the identities of all other peer nodes in the network, and therefore
they cannot coordinate this decision with those other nodes.  Finally,
replica nodes in peer-to-peer networks are not necessarily
homogeneous.  Some replica nodes may be very powerful with great
connectivity, whereas others may have limited inherent capacity to
handle content requests.

Previous load-balancing techniques in the literature base their
decisions on periodic or continuous updates containing information on
\emph{load} or \emph{available capacity}.  We refer to this
information as load-balancing information (LBI).  These techniques
have not been designed with peer-to-peer networks in mind and thus
\begin{itemize}
\item do not take into account the heterogeneity of peer nodes (e.g., \cite{genova00,mitzenmacher97}), or
\item  use techniques such as migration or handoff of tasks
that cannot be used in a peer-to-peer environment (e.g., \cite{lu96}), or
\item suffer from significant load oscillations, or ``herd behavior''
\cite{mitzenmacher97}, where peer nodes simultaneously forward an
unpredictable number of requests to replicas with low
reported load or high reported available capacity, causing them to
become overloaded.  This herd behavior defeats the attempt to provide
load-balancing.
\end{itemize}

Most of these techniques also depend on the timeliness of LBI updates.
The wide-area nature of peer-to-peer networks and the variation in
transfer delays among peer nodes makes guaranteeing the timeliness of
updates difficult.  Peer nodes will experience varying degrees of
staleness in the LBI updates they receive depending on their distance
from the source of updates.  Moreover, maintaining the timeliness of
LBI updates is also costly, since all updates must travel across the
Internet to reach interested peer nodes.  The smaller the inter-update
period and the larger the overlay peer network, the greater the
network traffic overhead incurred by LBI updates.  Therefore, in a
peer-to-peer environment, an effective load-balancing algorithm should
not be critically dependent on the timeliness of updates.

In this paper we propose a practical load-balancing algorithm,
Max-Cap, that makes decisions based on the inherent maximum capacities
of the replica nodes.  We define maximum capacity as the maximum
number of content requests per time unit that a replica claims it can
handle.  Alternative measures such as maximum (allowed) connections
can be used. The maximum capacity is like a contract by which the
replica agrees to abide. If the replica cannot sustain its advertised
rate, then it may choose to advertise a new maximum capacity.  Max-Cap
is not critically tied to the timeliness or frequency of LBI updates,
and as a result, when applied in a peer-to-peer environment,
outperforms algorithms based on load or available capacity, whose
benefits are heavily dependent on the timeliness of the updates.

We show that Max-Cap takes peer node heterogeneity into account unlike
algorithms based on load.  While algorithms based on available
capacity take heterogeneity into account, we show that they can suffer
from load oscillations in a peer-to-peer network in the presence of
small fluctuations in the workload even when the workload request rate
is well below the total maximum capacities of the replicas.  On the
other hand, Max-Cap avoids overloading replicas in such cases
and is more resilient to very large fluctuations in workload.  This is
because a key advantage of Max-Cap is that it uses information that is
not affected by changes in the workload.

Since it is most probable that each replica node will run other
applications besides the peer-to-peer content distribution
application, Max-Cap must also be able to handle fluctuations in
``extraneous load'' observed at the replicas.  This is load caused by
external factors such as other applications the users of the replica
node are running or network conditions occurring at the replica node.

We modify Max-Cap to perform load-balancing using the ``honored
maximum capacity'' of each replica.  This is the maximum capacity
minus the extraneous load observed at the replica.  Although the
honored maximum capacities may change frequently, the changes are
independent of fluctuations in the content request workload. As a
result, Max-Cap continues to provide better load-balancing than
availability-based algorithms even when there are large fluctuations
in the extraneous load.

In a peer-to-peer environment the expectation is that the set of
participating nodes changes constantly.  Since replica arrivals to and
departures from the peer network can affect the information carried in
LBI updates, we also compare Max-Cap against availability-based
algorithms when the set of replicas continuously changes.  We show
that Max-Cap is less affected by changes in the replica set than the
availability-based algorithms.

We evaluate load-based and availability-based algorithms and compare
them with Max-Cap in the context of CUP \cite{roussopoulos02}, a
protocol that asynchronously builds and maintains caches of index
entries in peer-to-peer networks through Controlled Update
Propagation.  The index entries for a particular content contain IP
addresses that point to replica nodes serving the content.
Load-balancing decisions are made from amongst these cached indices to
determine to which of the replica nodes a request for that content
should be forwarded.  CUP periodically propagates updates of desired
index entries down a conceptual tree (similar to an application-level
multicast tree) whose vertices are interested peer nodes.  We leverage
CUP's propagation mechanism by piggybacking LBI such as load or
available capacity onto the updates CUP propagates.

The rest of this paper is organized as follows.
Section~\ref{Architecture} briefly describes the CUP protocol and how
we use it to propagate the load-balancing information necessary to
implement the various load-balancing algorithms across replica nodes.
Section~\ref{Algorithms} introduces the algorithms compared.
Section~\ref{Experiments} presents experimental results showing that
in a peer-to-peer environment, Max-Cap outperforms the other
algorithms with much less or no overhead.
Section~\ref{LBRelatedWork} describes related work, and
Section~\ref{Conclusions} concludes the paper.

\section{CUP Protocol Design}
\label{Architecture}

In this section we briefly describe how we leverage the CUP protocol
to study the load-balancing problem in a peer-to-peer context.  CUP is
a protocol for maintaining caches of index entries in peer-to-peer
networks through \emph{C}ontrolled \emph{U}pdate \emph{P}ropagation.

CUP supports both structured and unstructured networks.  In structured
networks lookup queries for particular content follow a well-defined
path from the querying node toward an \emph{authority node}, which is
guaranteed to know the location of the content within the network.  In
unstructured networks lookup queries are flooded haphazardly
throughout the network until a node that knows the location of the
content is met.  In this paper, we will describe how CUP works within
structured networks \cite{ratnasamy01a,rowstron01b,stoica01,zhao01}.

In CUP every node in the peer-to-peer network maintains two logical
channels per neighbor: a query channel and an update channel.  The
query channel is used to forward lookup queries for content of
interest to the neighbor that is closest to the authority node for
that content.  The update channel is used to forward query responses
asynchronously to a neighbor.  These query responses contain sets of
index entries that point to nodes holding the content in question.
The update channel is also used to update the index entries that are
cached at the neighbor.

\begin{figure}
\centerline{\includegraphics[width=7cm, height=7cm]{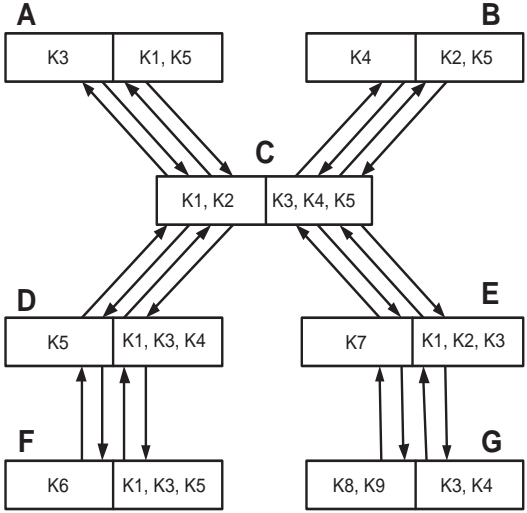}}
\caption[]{\small CUP Trees}
\label{fig:CUPTrees}
\end{figure}

Figure~\ref{fig:CUPTrees} shows a snapshot of CUP in progress in a
network of seven nodes.  The four logical channels are shown between
each pair of nodes. The left half of each node shows the set of
content items for which the node is the authority.  The right half
shows the set of content items for which the node has cached index
entries as a result of handling lookup queries.  For example, node A
is the authority node for content $K3$ and nodes C,D,E,F, and G have
cached index entries for content $K3$. The process of querying and
updating index entries for a particular content $K$ forms a CUP tree
whose root is the authority node for content $K$.  The branches of the
tree are formed by the paths traveled by lookup queries from other
nodes in the network. For example, in Figure~\ref{fig:CUPTrees}, node
A is the root of the CUP tree for $K3$ and branch \{F,D,C,A\} has
grown as a result of a lookup query for $K3$ at node F.

It is the authority node A for content $K3$ which is guaranteed to
know the location of all nodes, called \emph{content replica nodes} or
simply \emph{replicas}, that serve content $K3$.  Replica nodes first
send birth messages to authority A to indicate they are serving
content $K3$.  They may also send periodic refreshes or invalidation
messages to A to indicate they are still serving or no longer serving
the content.  A then forwards on any birth, refresh or invalidation
messages it receives, which are propagated down the CUP tree to all
interested nodes in the network.  For example, in
Figure~\ref{fig:CUPTrees} any update messages for index entries
associated with content $K3$ that arrive at A from replica nodes are
forwarded down the $K3$ CUP tree to C at level 1, D and E at level 2,
and F and G at level 3.

CUP has been extensively studied in \cite{roussopoulos02}.  While the
specific update propagation protocol CUP uses has been shown to
provide benefits such as greatly reducing the latency of lookup
queries, the specific CUP protocol semantics are not required for the
purposes of load-balancing.  We simply leverage the update propagation
mechanism of CUP to push LBI such as replica load or capacity to
interested peer nodes throughout the overlay network.  These peer
nodes can then use this information when choosing to which replica a
client request should be forwarded.

\section{The Algorithms}
\label{Algorithms}

We evaluate two different algorithms, Inv-Load and Avail-Cap.  Each is
representative of a different class of algorithms that have been
proposed in the distributed systems literature.  We study how these
algorithms perform when applied in a peer-to-peer context and compare
them with our proposed algorithm, Max-Cap.  These three algorithms
depend on different LBI being propagated, but their overall goal is
the same: to balance the demand for content fairly across the set of
replicas providing the content.  In particular, the algorithm should
avoid overloading some replicas while underloading others, especially
when the aggregate capacity of all replicas is enough to handle the
content request workload.  Moreover, the algorithm should prevent
individual replicas from oscillating between being overloaded and
underloaded.

Oscillation is undesirable for two reasons.  First, many applications
limit the number of requests a host can have outstanding.  This means
that when a replica node is overloaded, it will drop any request it
receives.  This forces the requesting client to resend its request
which has a negative impact on response time.  
Even for applications that allow
requests to be queued while a replica node is overloaded
the queueing delay incurred will also increase the average response
time.  Second, in a peer-to-peer network, the issue of fairness is
sensitive.  The owners of replica nodes are likely not to want their
nodes to be overloaded while other nodes in the network are
underloaded.  An algorithm that can fairly distribute the request
workload without causing replicas to oscillate between being
overloaded and underloaded is preferable.

We describe each of the algorithms we evaluate in turn:

\emph{Allocation Proportional to Inverse Load} (Inv-Load).  There are
many load-balancing algorithms that base the allocation decision on
the load observed at and reported by each of the serving entities (see
Related Work Section~\ref{LBRelatedWork}).  The representative
load-based algorithm we examine in this paper is Inv-Load, based on
the algorithm presented by Genova et al. \cite{genova00}.  In this
algorithm, each peer node in the network chooses to forward a request
to a replica with probability inversely proportional to the load
reported by the replica.  This means that the replica with the
smallest reported load (as of the last report received) will receive
the most requests from the node.  Load is defined as the number of
request arrivals at the replica per time unit.  Other possible load
metrics include the number of request connections open at the replica
at reporting time \cite{aversa00} or the request queue length at the
replica \cite{dahlin99}.

The Inv-Load algorithm has been shown to perform as well as or better
than other proposed algorithms in a homogeneous environment and for
this reason we focus on this algorithm in this study.  But, as we show
in Section~\ref{Heterogeneity}, Inv-Load does not handle node
heterogeneity well.

As we will see in Section~\ref{Inv-Load-Heterogeneity}, Inv-Load is not
designed to handle replica node heterogeneity.

\emph{Allocation Proportional to Available Capacity} (Avail-Cap).  In
this algorithm, each peer node chooses to forward a request to a
replica with probability proportional to the available capacity
reported by the replica.  Available capacity is the maximum request
rate a replica can handle minus the load (actual request rate)
experienced at the replica.  This algorithm is based on the algorithm
proposed by Zhu et al. \cite{zhu98} for load sharing in a cluster of
heterogeneous servers.  Avail-Cap takes into account heterogeneity
because it distinguishes between nodes that experience the same load
but have different maximum capacities.

Intuitively, Avail-Cap seems like it should work; it handles
heterogeneity by sending more requests to the replicas that are
currently more capable. Replicas that are overloaded report an
available capacity of zero and are excluded from the allocation
decision until they once more report a positive available capacity.
Unfortunately, as we will show in Section ~\ref{Oscillation}, this
exclusion can cause Avail-Cap to suffer from wild load oscillations.

Both Inv-Load and Avail-Cap implicitly assume that the load or
available capacity reported by a replica remains roughly constant
until the next report.  Since both these metrics are directly affected
by changes in the request workload, both algorithms require that
replicas periodically update their LBI.  (We assume replicas are not
synchronized in when they send reports.)  Decreasing the period
between two consecutive LBI updates increases the timeliness of the
LBI at a cost of higher overhead (in number of updates pushed through
the peer-to-peer network).  In large peer-to-peer networks, there may
be several levels in the CUP tree down which updates will have to
travel, and the time to do so could be on the order of seconds.
 
\emph{Allocation Proportional to Maximum Capacity} (Max-Cap).  This is
the algorithm we propose.  In this algorithm, each peer node chooses
to forward a request to a replica with probability proportional to the
maximum capacity of the replica.  The maximum capacity is a contract
each replica advertises indicating the number of requests the replica
claims to handle per time unit.  Unlike load and available capacity,
the maximum capacity of a replica is not affected by changes in the
content request workload.  Therefore, Max-Cap does not depend on the
timeliness of the LBI updates.  In fact, replicas only push updates
down the CUP tree when they choose to advertise a new maximum
capacity.  This choice depends on extraneous factors that are
unrelated to and independent of the workload (see
Section~\ref{ExtraneousLoad}).  If replicas rarely choose to change
contracts, Max-Cap incurs near-zero overhead.  We believe that this
independence of the timeliness and frequency of updates makes Max-Cap
practical and elegant for use in peer-to-peer networks.

\section{Experiments}
\label{Experiments}

In this section we describe experiments that measure the ability of
the Inv-Load, Avail-Cap and Max-Cap algorithms to balance requests for
content fairly across the replicas holding the content.  We simulate a
content-addressable network (CAN) \cite{ratnasamy01a} using the
Stanford Narses simulator \cite{maniatis01}. A CAN is an example of a
structured peer-to-peer network, defined in
Section~\ref{Architecture}. In each of these experiments, requests for
a specific piece of content are posted at nodes throughout the CAN
network for 3000 seconds.  Using the CUP protocol described in
Section~\ref{Architecture}, a node that receives a content request
from a local client retrieves a set of index entries pointing to
replica nodes that serve the content.  The node applies a
load-balancing algorithm to choose one of the replica nodes.  It then
points the local client making the content request at the chosen
replica.  

The simulation input parameters include: the number of nodes in the
overlay peer-to-peer network, the number of replica nodes holding the
content of interest, the maximum capacities of the replica nodes, the
distribution of content request inter-arrival times, a seed to feed
the random number generators that drive the content request arrivals
and the allocation decisions of the individual nodes, and the LBI
update period, which is the amount of time each replica waits before
sending the next LBI update for the Inv-Load and Avail-Cap algorithms.

We assign maximum capacities to replica nodes by applying results from
recent work that measures the upload capabilities of nodes in Gnutella
networks \cite{saroiu02}. This work has found that for the Gnutella
network measured, around 10\% of nodes are connected through dial-up
modems, 60\% are connected through broadband connections such as cable
modem or DSL where the upload speed is about ten times that of dial-up
modems, and the remaining 30\% have high-end connections with upload
speed at least 100 times that of dial-up modems.  Therefore we assign
maximum capacities of 1, 10, and 100 requests per second to nodes with
probabilty of 0.1, 0.6, and 0.3, respectively.

In all the experiments we present in this paper, the number of nodes
in the network is 1024, each individually deciding how to distribute
its incoming content requests across the replica nodes.  We use both
Poisson and Pareto request inter-arrival distributions, both of which
have been found to hold in peer-to-peer networks
\cite{cao02,markatos02}.

We present five experiments.  First we show that Inv-Load cannot
handle heterogeneity.  We then show that while Avail-Cap takes replica
heterogeneity into account, it can suffer from significant load
oscillations caused by even small fluctuations in the workload.  We
compare Max-Cap with Avail-Cap for both Poisson and bursty Pareto
arrivals.  We also compare the effect on the performances of Avail-Cap
and Max-Cap when replicas continuously enter and leave the system.
Finally, we study the effect on Max-Cap when replicas cannot always
honor their advertised maximum capacities because of significant
extraneous load.

\subsection{Inv-Load and Heterogeneity}
\label{Inv-Load-Heterogeneity}

\begin{figure}
\centerline{\includegraphics[width=8cm]{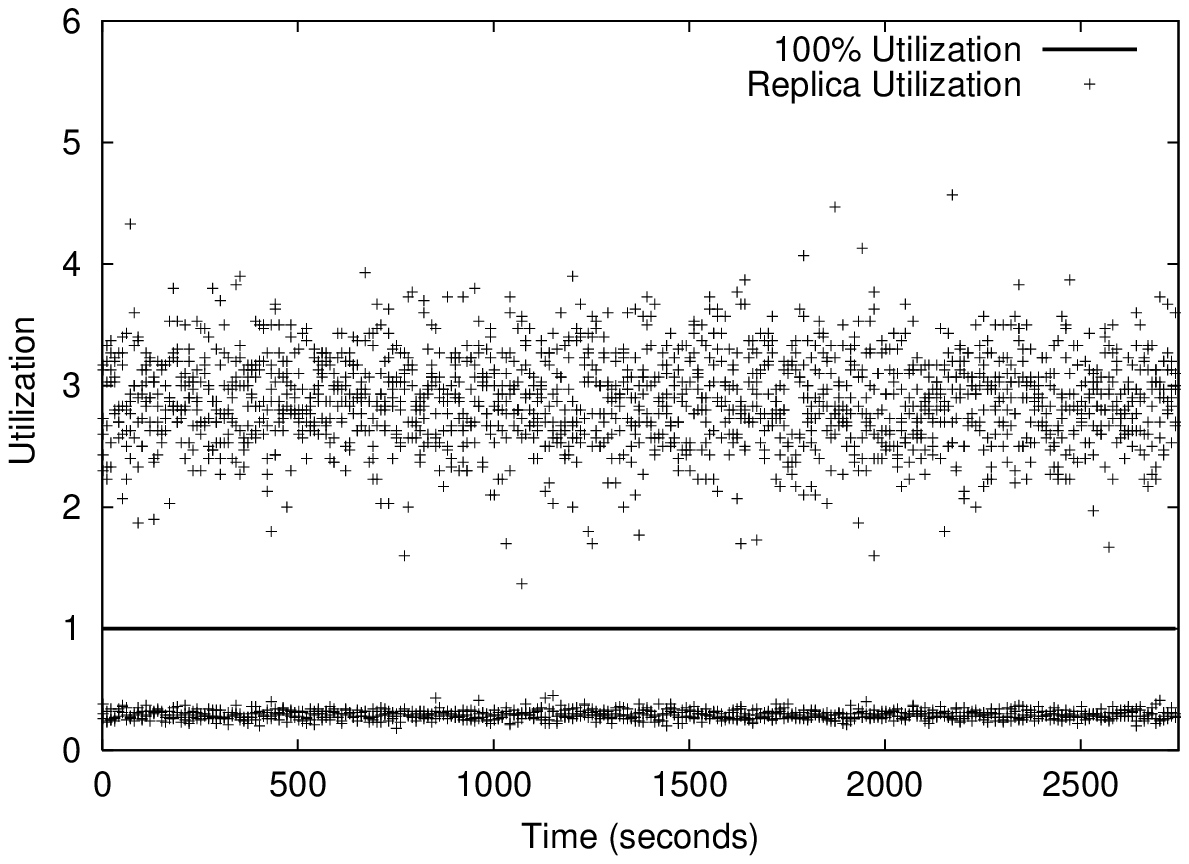}}
\caption{\small Replica Utilization versus Time for InvLoad with heterogeneous replicas.}
\label{Inv-Load-Hetero-289-Scatterplot}
\end{figure}

In this experiment, we examine the performance of Inv-Load in a
heterogeneous peer-to-peer environment.  We use a fairly short
inter-update period of one second, which is quite aggressive in a
large peer-to-peer network.  We have ten replica nodes that serve the
content item of interest, and we generate request rates for that item
according to a Poisson process with an arrival rate that is 80\% of
the total maximum capacities of the replicas.  Under such a workload,
a good load-balancing algorithm should be able to avoid overloading
some replicas while underloading others.
Figure~\ref{Inv-Load-Hetero-289-Scatterplot} shows a scatterplot of
how the utilization of each replica proceeds with time when using
Inv-Load.  We define utilization as the request arrival rate observed
by the replica divided by the maximum capacity of the replica.  In
this graph, we do not distinguish among points of different replicas.
We see that throughout the simulation at any point in time, some
replicas are severely overutilized (over 250\%) while others are
lightly underutilized (around 25\%).

\begin{figure}
\centerline{\includegraphics[width=8cm]{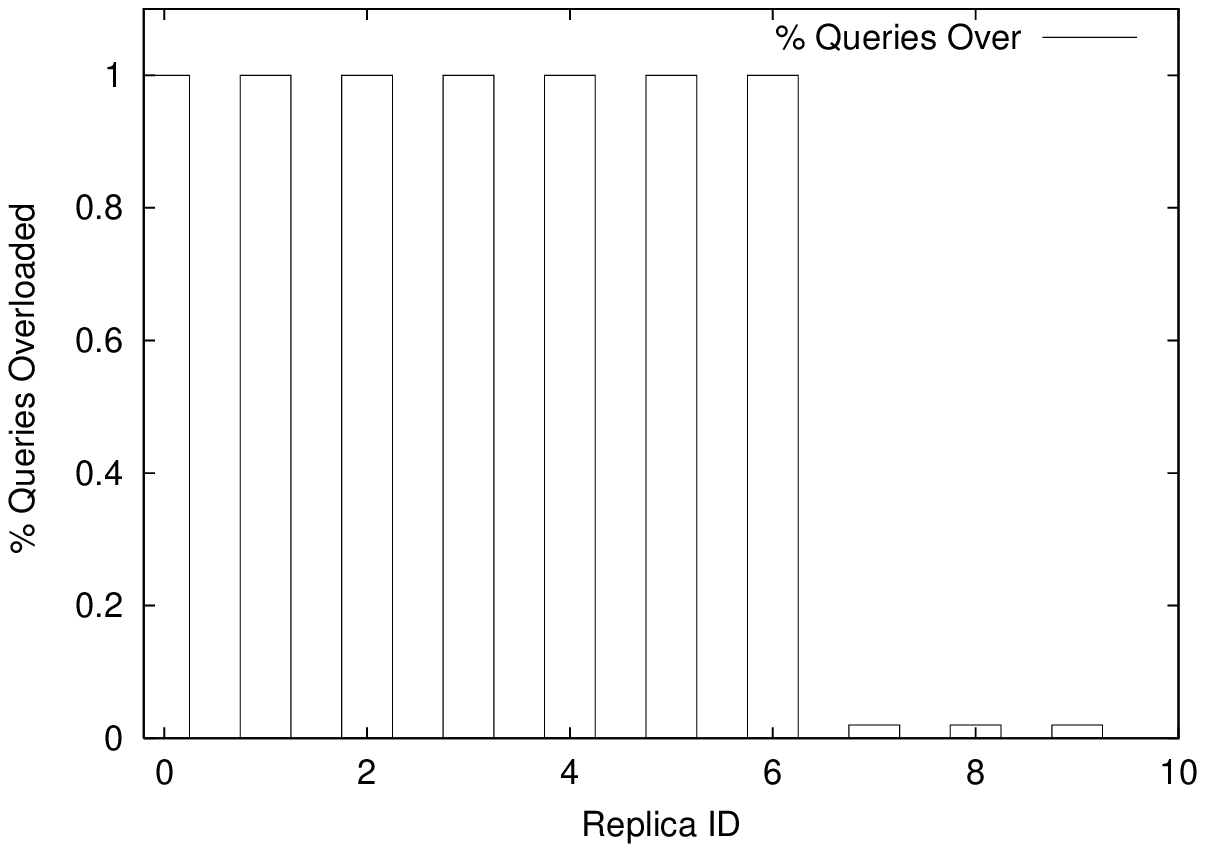}}
\caption{\small Percentage Overloaded Queries versus Replica ID for Inv-Load with heterogeneous replicas.}
\label{Inv-Load-Hetero-289-OverCapQueries}
\end{figure}

Figure~\ref{Inv-Load-Hetero-289-OverCapQueries} shows for each
replica, the percentage of all received requests that arrive while the
replica is overloaded.  This measurement gives a true picture of how
well a load-balancing algorithm works for each replica.  In
Figure~\ref{Inv-Load-Hetero-Scatterplot-AND-OverCapQueries}b, the
replicas that receive almost 100\% of their requests while overloaded
(i.e., replicas 0-6) are the low and middle-end replicas.  The
replicas that receive almost no requests while overloaded (i.e.,
replicas 7-9) are the high-end replicas.  We see that Inv-Load
penalizes the less capable replicas while giving the high-end replicas
an easy time.

Inv-Load is designed to perform well in a homogeneous environment.
When applied in a heterogeneous environment such as a peer-to-peer
network, it fails.  As we will see in the next section Max-Cap is much
better suited.  Apart from showing that Max-Cap has comparable load
balancing capability with no overhead in a homogeneous environment
(see Appendix), we do not consider Inv-Load in the remaining
experiments as our focus here is on heterogeneous environments.

\subsection{Avail-Cap versus Max-Cap}
\label{Oscillation}

In this set of experiments we examine the performance of Avail-Cap and
compare it with Max-Cap.  

\subsubsection{Poisson Request Arrivals}
\begin{figure}
\centerline{\includegraphics[width=8cm]{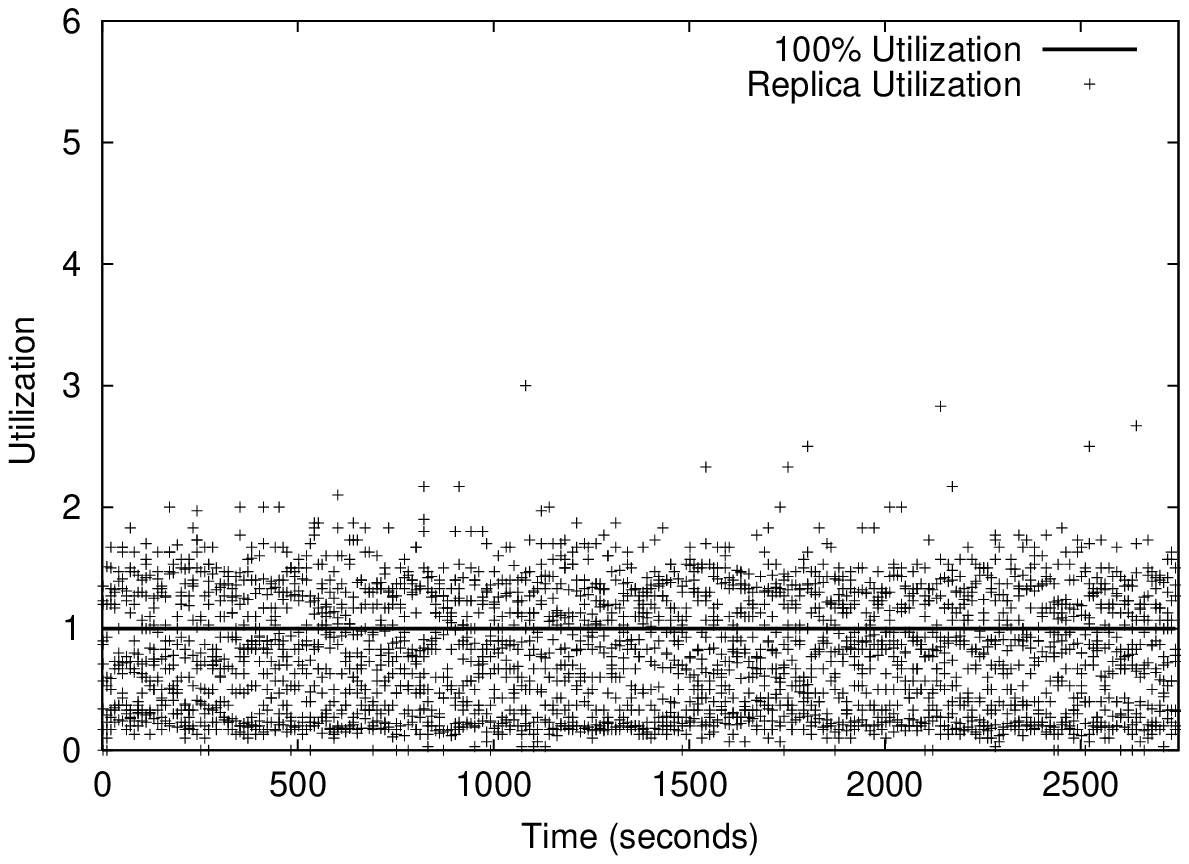}}
\caption{\small Replica Utilization versus Time for Avail-Cap with heterogeneous replicas.}
\label{Avail-Cap-289-ScatterPlot-pd-1}
\end{figure}

\begin{figure}
\centerline{\includegraphics[width=8cm]{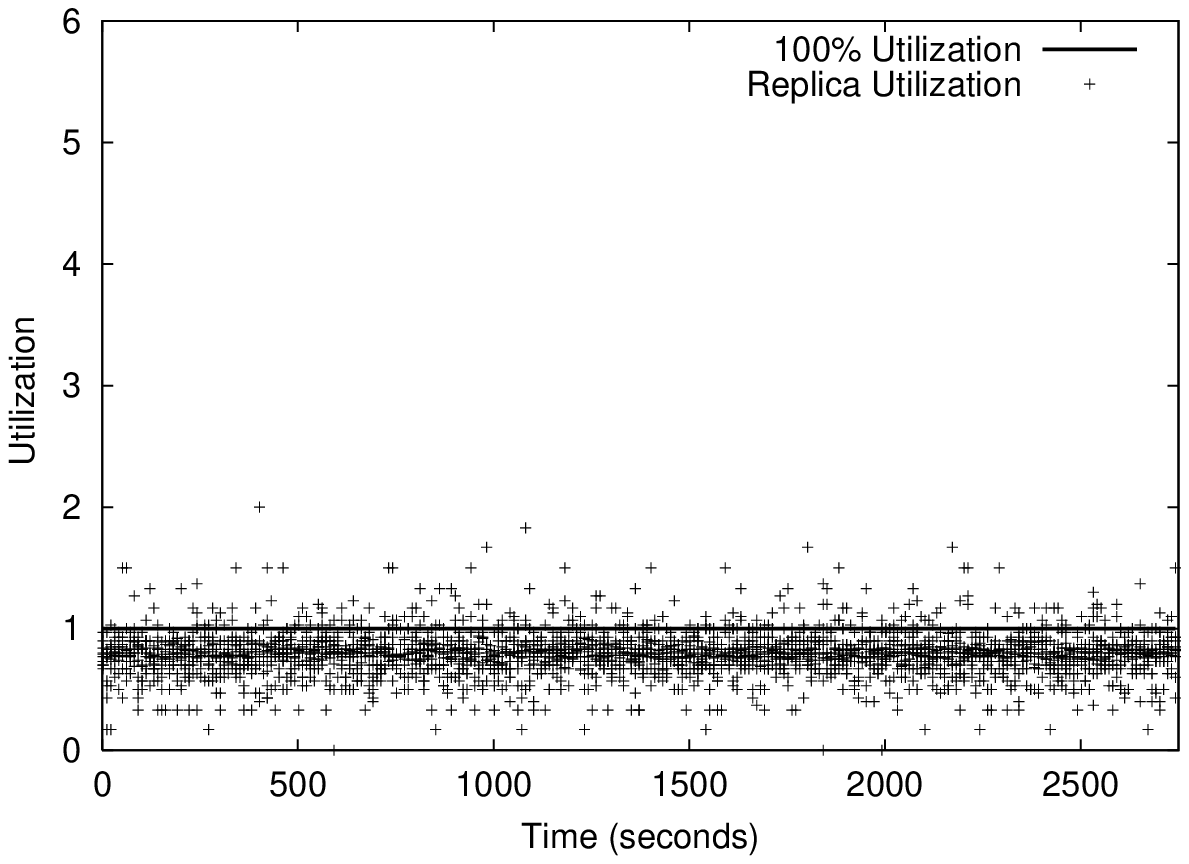}}
\caption{\small Replica Utilization v. Time for Max-Cap with heterogeneous replicas.}
\label{Max-Cap-289-Scatterplot}
\end{figure}

In Figures~\ref{Avail-Cap-289-ScatterPlot-pd-1} and
\ref{Max-Cap-289-Scatterplot} we show the replica utilization versus
time for an experiment with ten replicas with a Poisson request
arrival rate of 80\% the total maximum capacities of the replicas. For
Avail-Cap, we use an inter-update period of one second.  For Max-Cap,
this parameter is inapplicable since replica nodes do not send updates
unless they experience extraneous load (see
Section~\ref{ExtraneousLoad}).  We see that Avail-Cap consistently
overloads some replicas while underloading others.  In contrast,
Max-Cap tends to cluster replica utilization at around 80\%.  We ran
this experiment with a range of Poisson lambda rates and found similar
results for rates that were 60-100\% the total maximum capacities of
the replicas.  Avail-Cap consistently overloads some replicas while
underloading others whereas Max-Cap clusters replica utilization at
around X\% utilization, where X is the overall request rate divided by
the total maximum capacities of the replicas.

It turns out that in Avail-Cap, unlike Inv-Load, it is not the same
replicas that are consistently overloaded or underloaded throughout
the experiment.  Instead, from one instant to the next, individual
replicas oscillate between being overloaded and severely underloaded.

\begin{figure}
\centerline{\includegraphics[width=8cm]{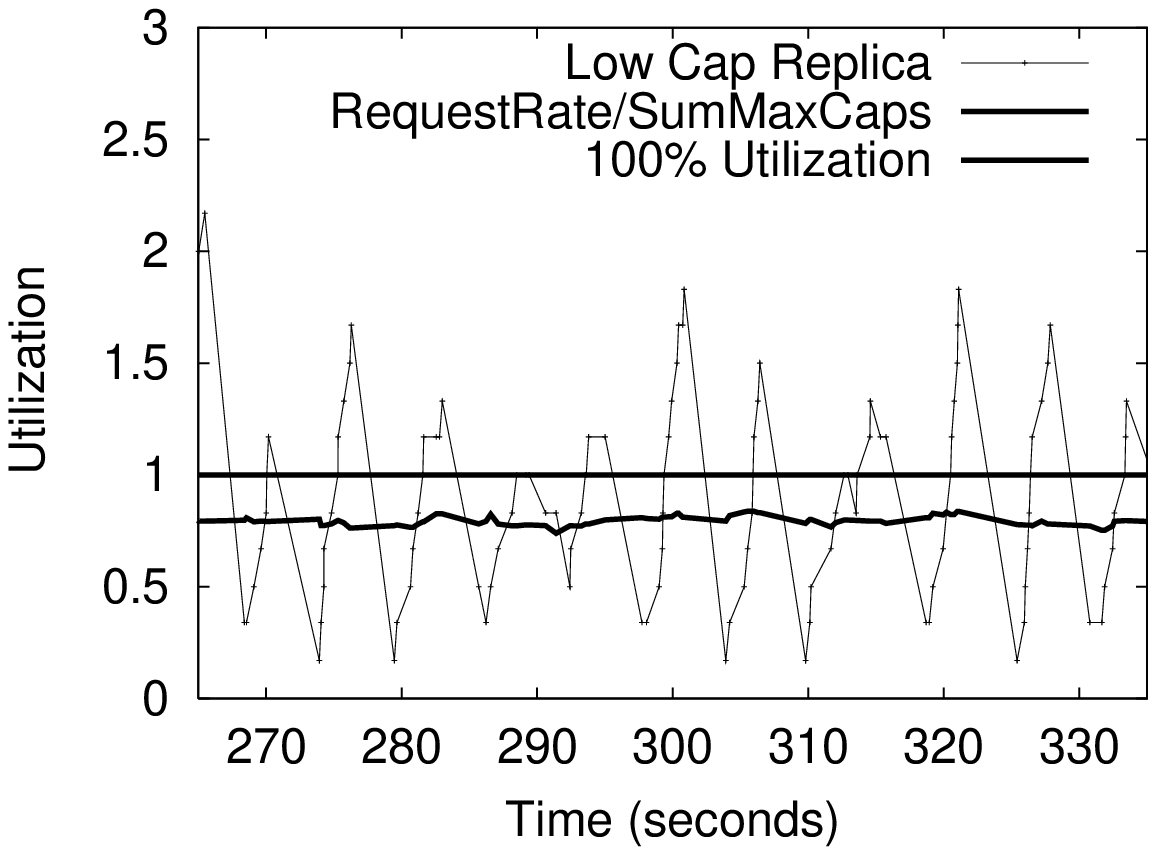}}
\caption{\small Low-end Replica Utilization versus Time for Avail-Cap, Poisson arrivals.}
\label{Avail-Cap-Low-Rep}
\end{figure}

\begin{figure}
\centerline{\includegraphics[width=8cm]{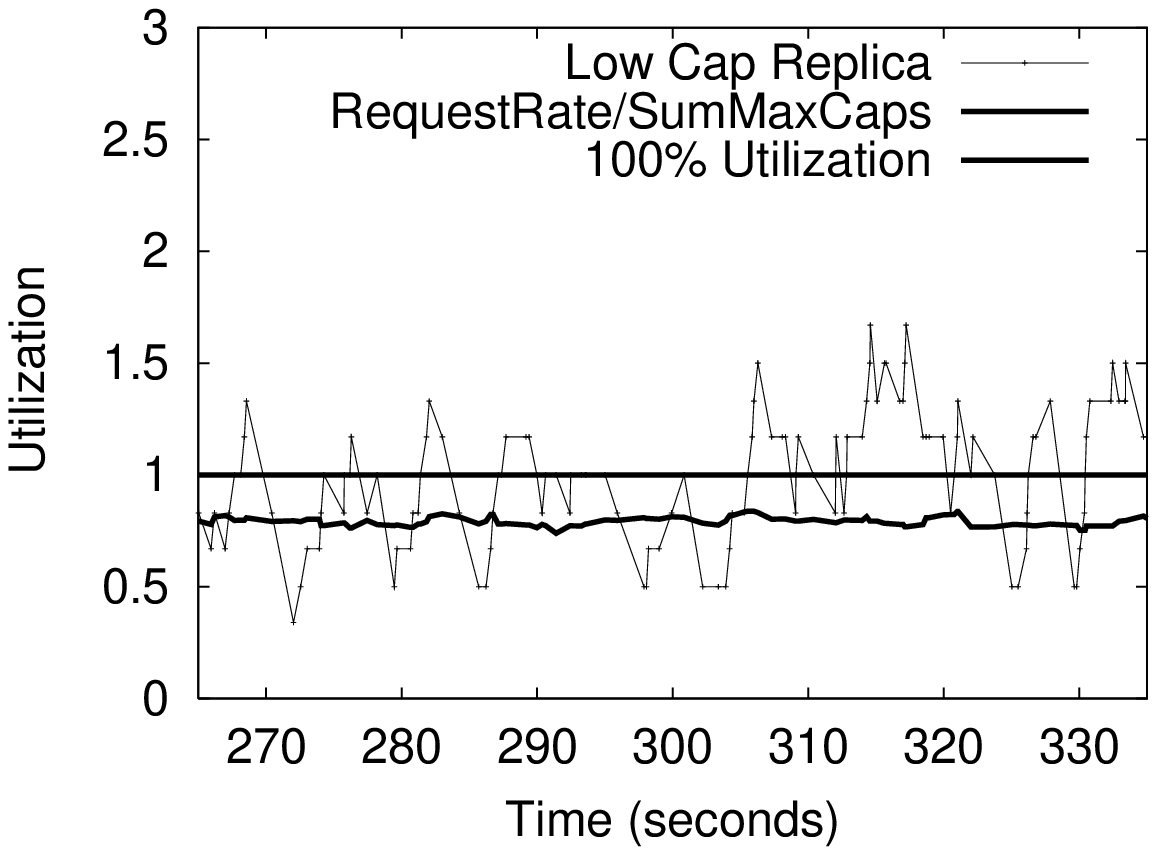}}
\caption{\small Low-end Replica Utilization versus Time for Max-Cap, Poisson arrivals.}
\label{Max-Cap-Low-Rep}
\end{figure}

\begin{figure}
\centerline{\includegraphics[width=8cm]{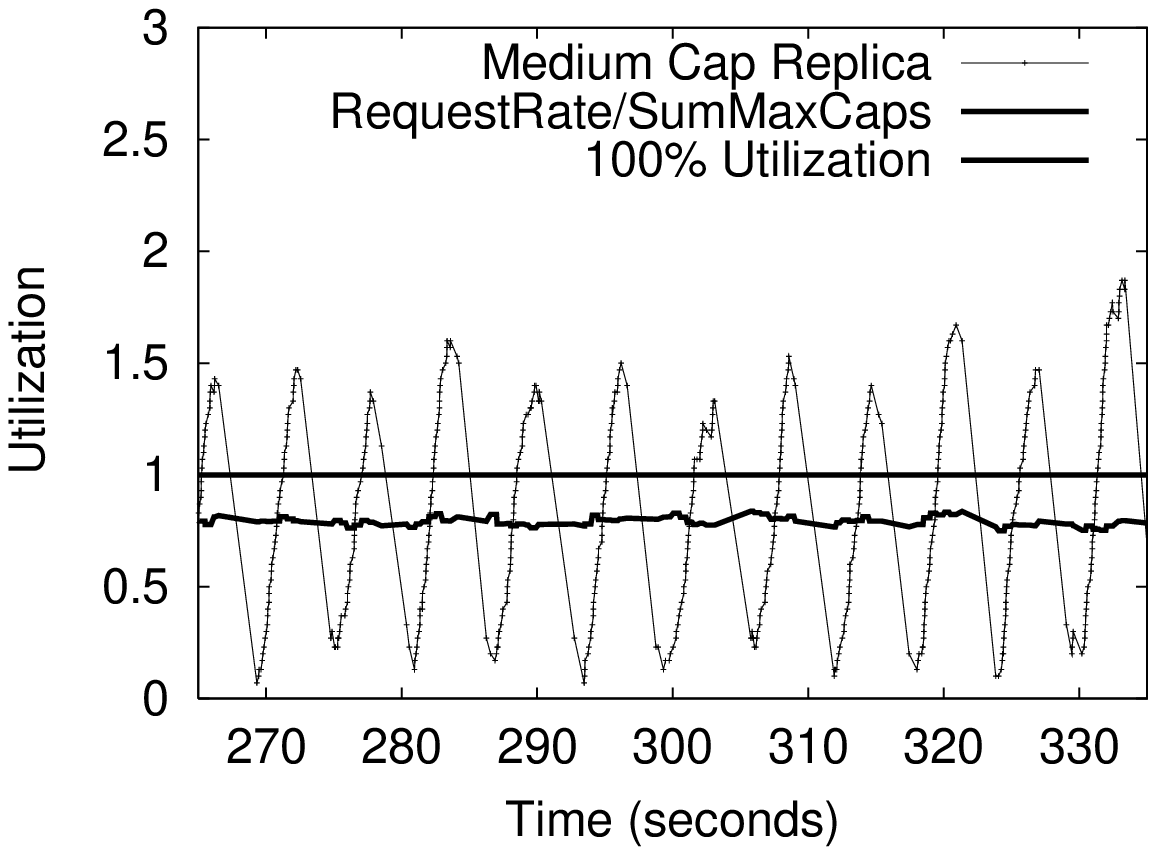}}
\caption{\small Medium-end Replica Utilization versus Time for Avail-Cap, Poisson arrivals.}
\label{Avail-Cap-Medium-Rep}
\end{figure}

\begin{figure}
\centerline{\includegraphics[width=8cm]{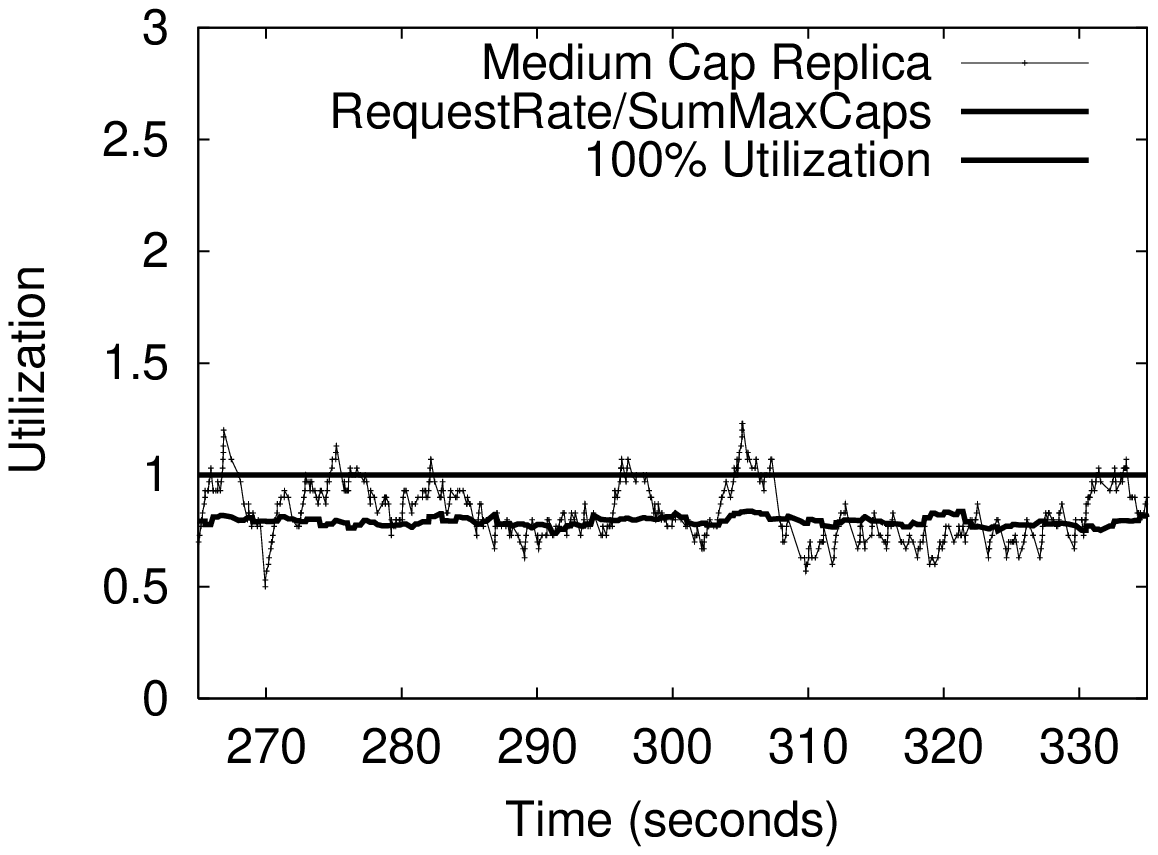}}
\caption{\small Medium-end Replica Utilization versus Time for Max-Cap, Poisson arrivals.}
\label{Max-Cap-Medium-Rep}
\end{figure}

\begin{figure}
\centerline{\includegraphics[width=8cm]{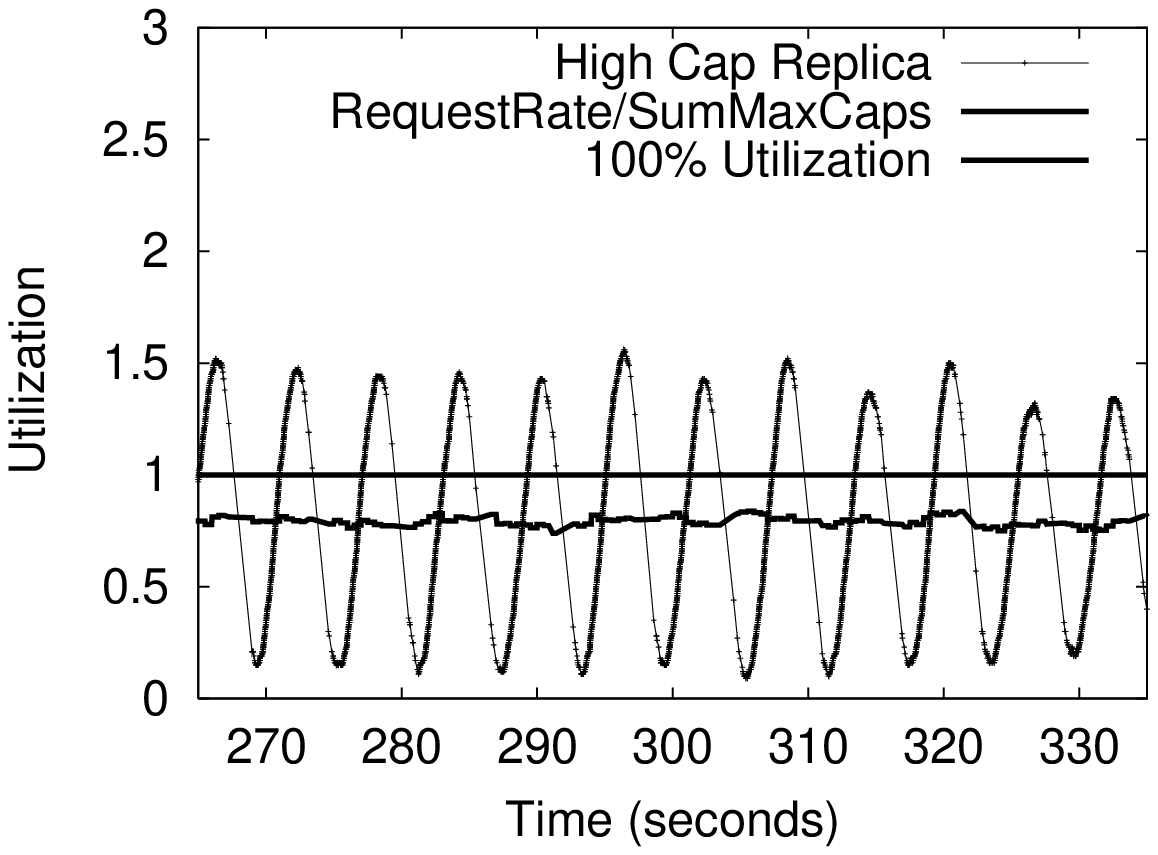}}
\caption{\small High-end Replica Utilization versus Time for Avail-Cap, Poisson arrivals.}
\label{Avail-Cap-High-Rep}
\end{figure}

\begin{figure}
\centerline{\includegraphics[width=8cm]{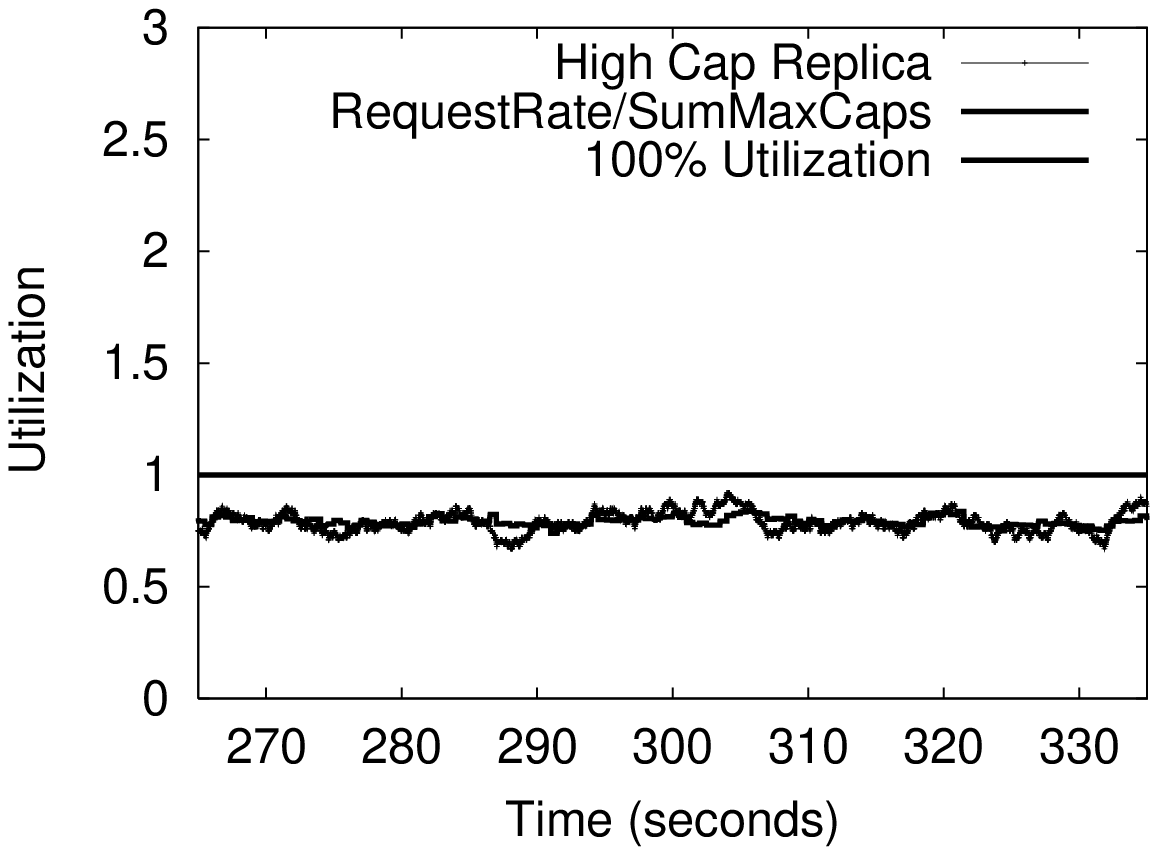}}
\caption{\small High-end Replica Utilization versus Time for Max-Cap, Poisson arrivals.}
\label{Max-Cap-High-Rep}
\end{figure}

We can see a sampling of this oscillation by looking at the
utilizations of some individual replicas over time.  In
Figures~\ref{Avail-Cap-Low-Rep}-\ref{Max-Cap-High-Rep}, we plot the
utilization over a one minute period in the experiment for a
representative replica from each of the replica classes (low, medium,
and high maximum capacity).  We also plot the ratio of the overall
request rate to the total maximum capacities of the replicas and the
line $y=1$ showing 100\% utilization. We see that for all replica
classes, Avail-Cap suffers from significant oscillation when compared
with Max-Cap which causes little or no oscillation.  This behavior
occurs throughout the experiment.

\begin{figure}
\centerline{\includegraphics[width=8cm]{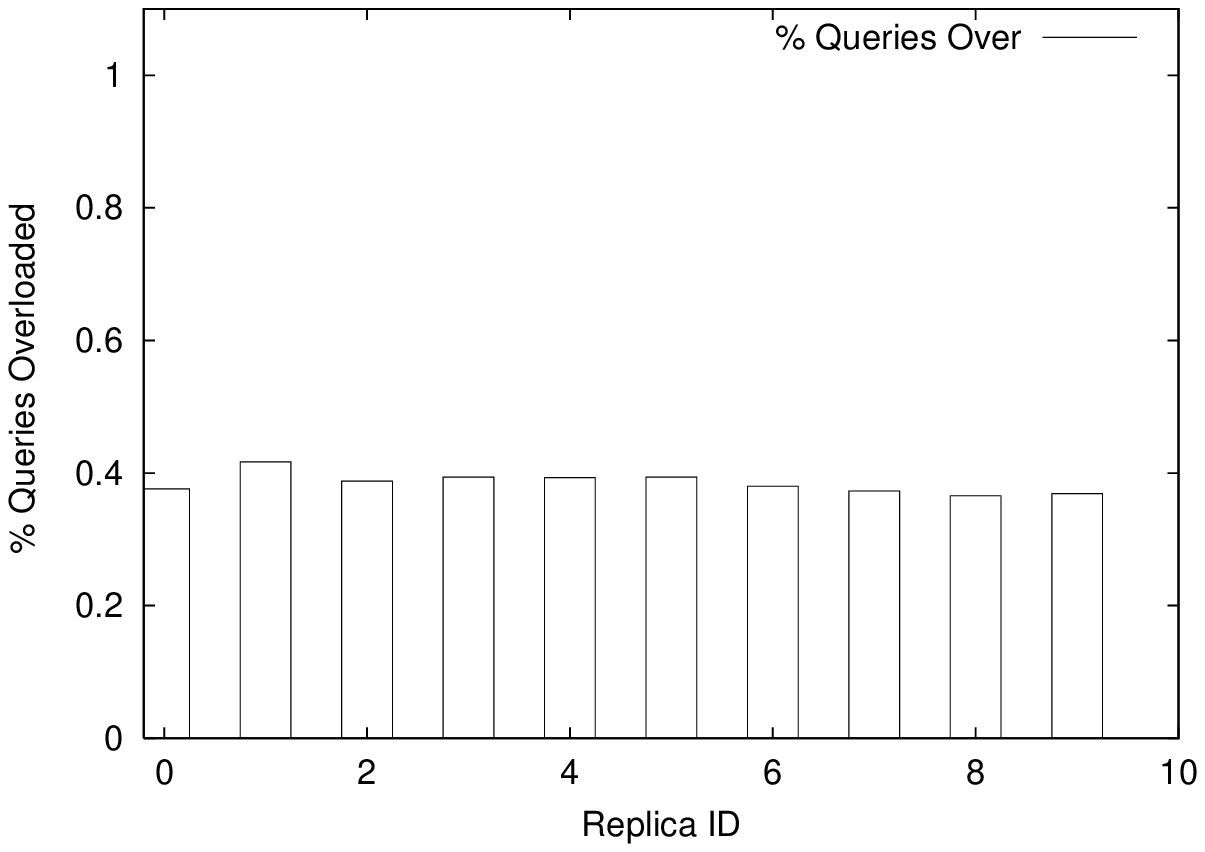}}
\caption{\small Percentage Overloaded Queries versus Replica ID for Avail-Cap, with inter-update period of 1 second.}
\label{Avail-Cap-289-OverCapQueries-pd-1}
\end{figure}

\begin{figure}
\centerline{\includegraphics[width=8cm]{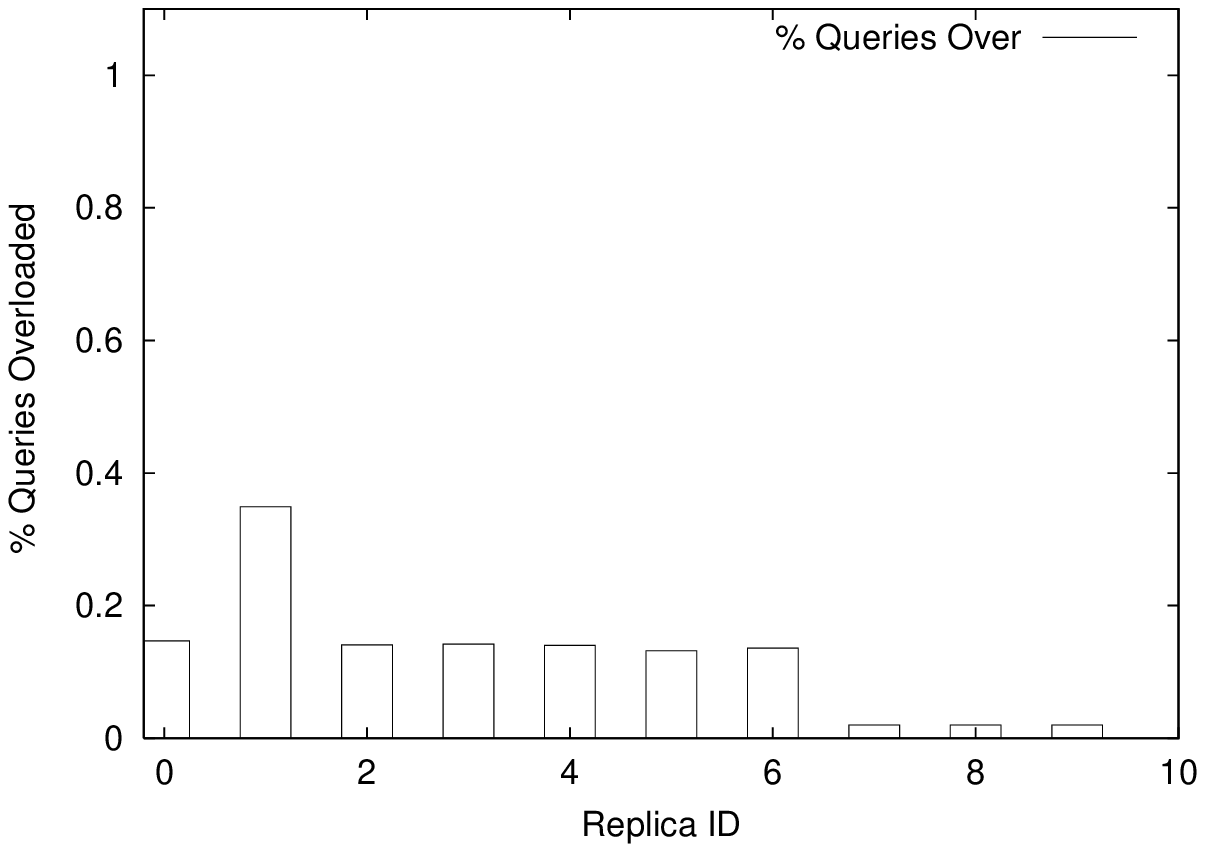}}
\caption{\small Percentage Overloaded Queries versus Replica ID for Max-Cap.}
\label{Max-Cap-289-OverCapQueries}
\end{figure}

Figures~\ref{Avail-Cap-289-OverCapQueries-pd-1} and
\ref{Max-Cap-289-OverCapQueries} show the percentage of requests that
arrive at each replica while the replica is overloaded for Avail-Cap
and Max-Cap respectively.  We see that Max-Cap achieves much lower
percentages than Avail-Cap.

\begin{figure}
\centerline{\includegraphics[width=8cm]{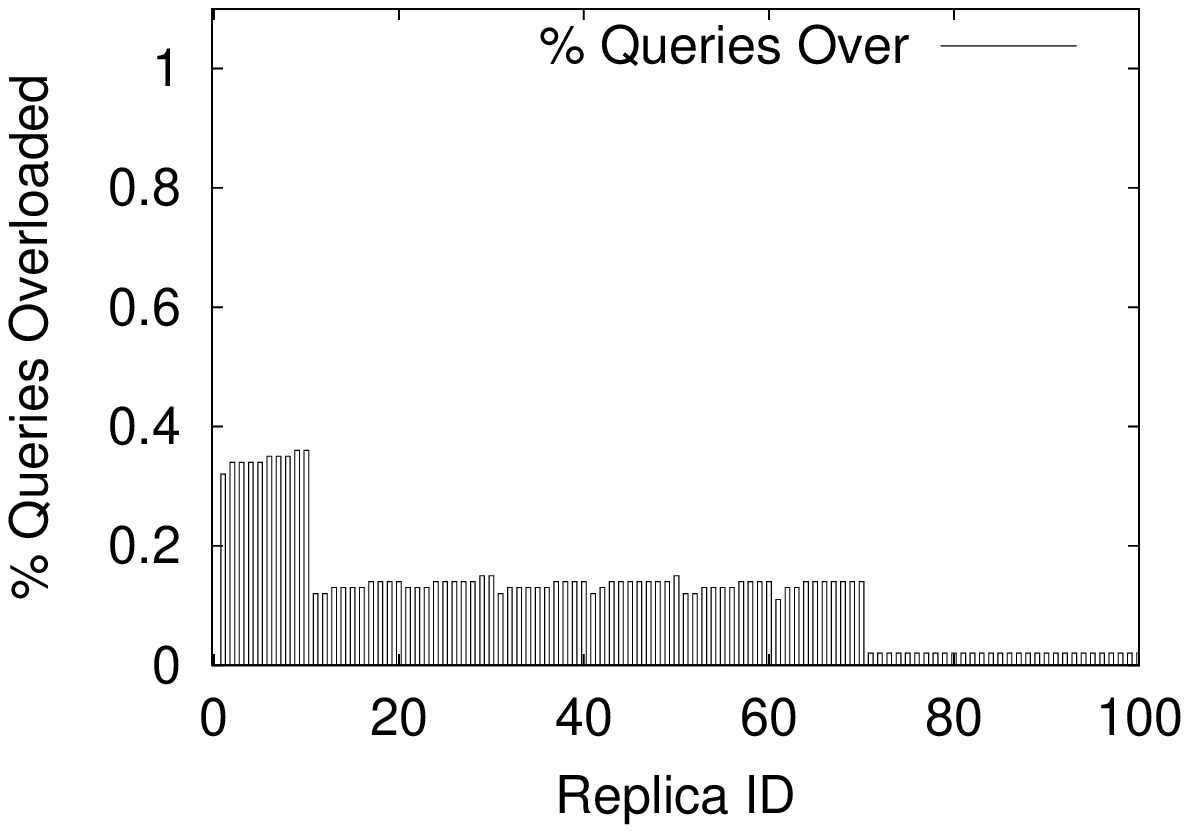}}
\caption{\small Percentage Overloaded Queries versus Replica ID for Max-Cap for ten experiments.}
\label{Max-Cap-289-OverCapQueries-ManySeeds}
\end{figure}

We also see in Figure~\ref{Max-Cap-289-OverCapQueries} that Max-Cap
exhibits a step-like behavior where the low-capacity replica (replica
1) is overloaded for about 35\% of its queries, the middle-capacity
replicas (replicas 0 and 2-6) are each overloaded for about 14\% of
their queries, and the high-capacity replicas (replicas 7-9) are each
overloaded for about 0.1\% of their queries.  To verify that this step
effect is not a random coincidence, we ran a series of experiments,
with ten replicas per experiment, and Poisson arrivals of 80\% the
total maximum capacity, each time varying the seed fed to the
simulator.  In Figure~\ref{Max-Cap-289-OverCapQueries-ManySeeds}, we
show the overloaded percentages for ten of these experiments.  On the
x-axis we order replicas according to maximum capacity, with the
low-capacity replicas plotted first (replica IDs 1 through 10),
followed by the middle-capacity replicas (replica IDs 11-70), followed
by the high-capacity replicas (replica IDs 71-100).  From the figure
we see that the step behavior consistently occurs.  This step behavior
occurs because the lower-capacity replicas have less tolerance for
noise in the random coin tosses the nodes perform while assigning
requests.  They also have less tolerance for small fluctuations in the
request rate.  As a result, lower-capacity replicas are overloaded
more easily than higher-capacity replicas.

\begin{figure}
\centerline{\includegraphics[width=8cm]{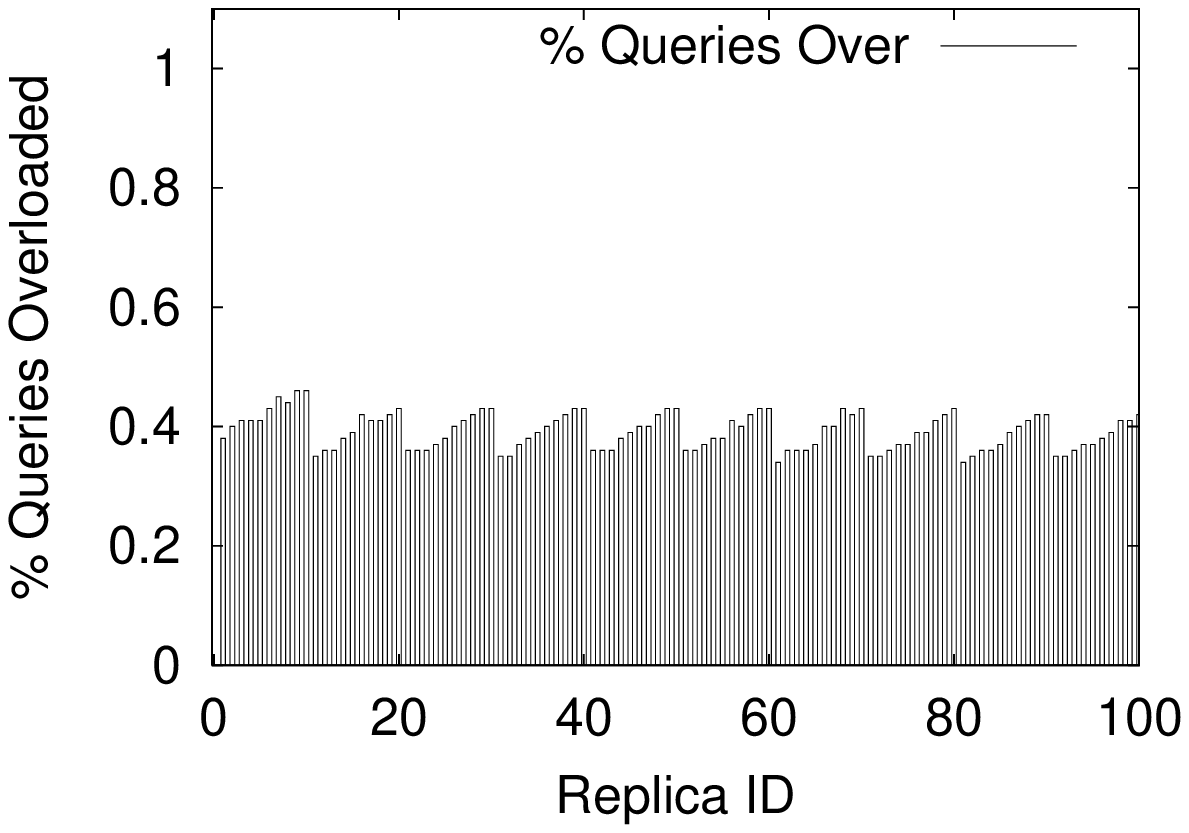}}
\caption{\small Percentage Overload Queries versus Replica ID for Avail-Cap with inter-update period of 1 second, for ten experiments.}
\label{Avail-Cap-289-OverCapQueries-ManySeeds-pd-1}
\end{figure}

Figure~\ref{Avail-Cap-289-OverCapQueries-pd-1} shows that Avail-Cap
with an inter-update period of one second causes much higher
percentages than Max-Cap (more than twice as high for the medium and
high-end replicas).  Avail-Cap also causes fairly even overloaded
percentages at around 40\%.  Again, to verify this evenness, in
Figure~\ref{Avail-Cap-289-OverCapQueries-ManySeeds-pd-1}, we show for
a series of ten experiments, the percentage of requests that arrive at
each replica while the replica is overloaded.  We see that Avail-Cap
consistently achieves roughly even percentages (at around 40\%) across
all replica types in contrast to the step effect observed by Max-Cap.
This can be explained by looking at the oscillations observed by
replicas in Figures~\ref{Avail-Cap-Low-Rep}-\ref{Max-Cap-High-Rep}.
In Avail-Cap, each replica
is overloaded for roughly the same amount of time regardless of
whether it is a low, medium or high-capacity replica.  This means that
while each replica is getting the correct proportion of requests, it
is receiving them at the wrong time and as a result all the replicas
experience roughly the same overloaded percentages.  In Max-Cap, we
see that replicas with lower maximum capacity are overloaded for more
time that higher-capacity replicas.
Consequently, higher-capacity replicas tend to have smaller overload
percentages than lower-capacity replicas.

\begin{figure}
\centerline{\includegraphics[width=8cm]{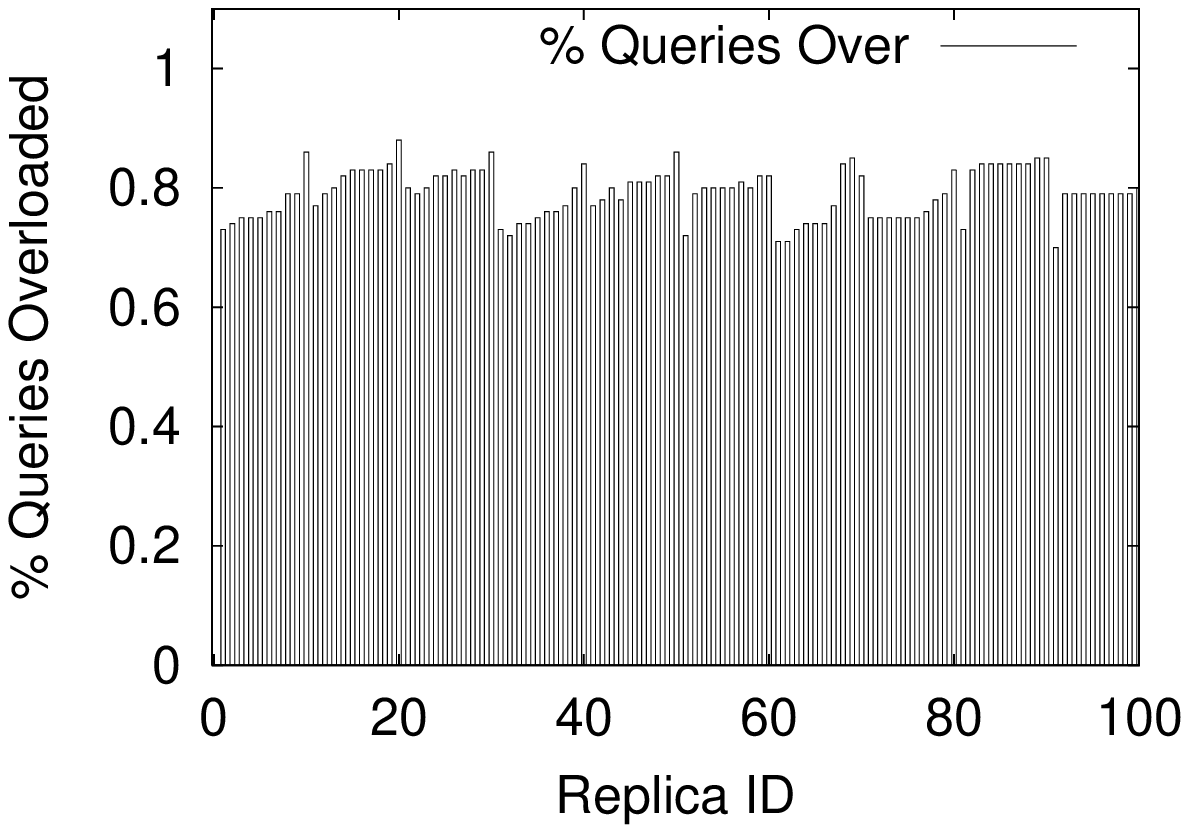}}
\caption{\small Percentage Overload Queries versus Replica ID for Avail-Cap with inter-update period of 10 seconds, for ten experiments.}
\label{Avail-Cap-289-OverCapQueries-ManySeeds-pd-10}
\end{figure}

The performance of Avail-Cap is highly dependent on the inter-update
period used.  We find that as we increase the period and available
capacity updates grow more stale, the performance of Avail-Cap suffers
more.  As an example, in
Figure~\ref{Avail-Cap-289-OverCapQueries-ManySeeds-pd-10}, we show the
overloaded query percentages in the same series of ten experiments for
Avail-Cap with a period of ten seconds.  The overloaded percentages
jump up to about 80\% across the replicas.

In a peer-to-peer environment, we argue that Max-Cap is a more practical
choice than Avail-Cap.  First, Max-Cap typically incurs no overhead.
Second, Max-Cap can handle workload rates that are below 100\% the total
maximum capacities and can handle small fluctuations in the workload
as are typical in Poisson arrivals.  

A question remaining is how do Avail-Cap and Max-Cap compare when
workload rates fluctuate beyond the total maximum capacities of the
replicas?  Such a scenario can occur for example when requests are
bursty, as when inter-request arrival times follow a Pareto
distribution.  We examine Pareto arrivals next.

\subsubsection{Pareto Request Arrivals}

Recent work has observed that in some peer-to-peer networks, request
inter-arrivals exhibit burstiness on several time scales
\cite{markatos02}, making the Pareto distribution a good candidate for
modeling these inter-arrival times.  

The Pareto distribution has two parameters associated with it: the
shape parameter $\alpha > 0$ and the scale parameter $\kappa > 0$.
The cumulative distribution function of inter-arrival time durations
is $F(x) = 1-({\frac {\kappa}{(x+\kappa)}})^\alpha.$ This distribution
is heavy-tailed with unbounded variance when $\alpha < 2$.  For
$\alpha > 1$, the average number of query arrivals per time unit is
equal to $\frac {(\alpha -1)}{\kappa}$.  For $\alpha <= 1$, the
expectation of an inter-arrival duration is unbounded and therefore
the average number of query arrivals per time unit is 0.

Typically, Pareto request arrivals are characterized by frequent and
intense bursts of requests followed by idle periods of varying
lengths.  During the bursts, the average request arrival rate can be
many times the total maximum capacities of the replicas.  We
present a representative experiment in which $\alpha$ and $\kappa$ are
1.1 and 0.000346 respectively.  These particular settings cause bursts
of up to 230\% the total maximum capacities of the replicas.
With such intense bursts, no load-balancing algorithm can be expected
to keep replicas underloaded.  Instead the best an algorithm can do is
to have the oscillation observed by each replica's utilization match
the oscillation of the ratio of overall request rate to total maximum
capacities.

\begin{figure}
\centerline{\includegraphics[width=8cm]{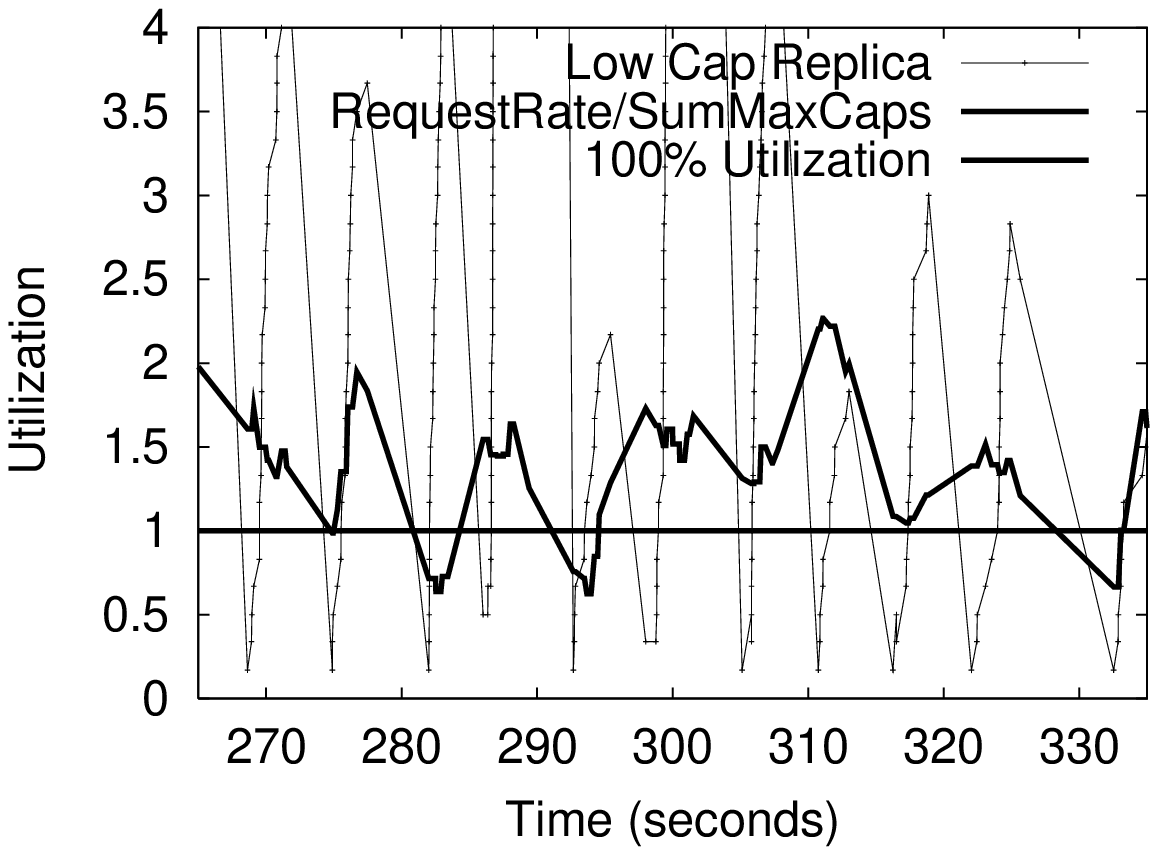}}
\caption{\small Low-capacity Replica Utilization versus Time for Avail-Cap, Pareto arrivals.}
\label{Pareto-Repl1-Avail-Cap}
\end{figure}

\begin{figure}
\centerline{\includegraphics[width=8cm]{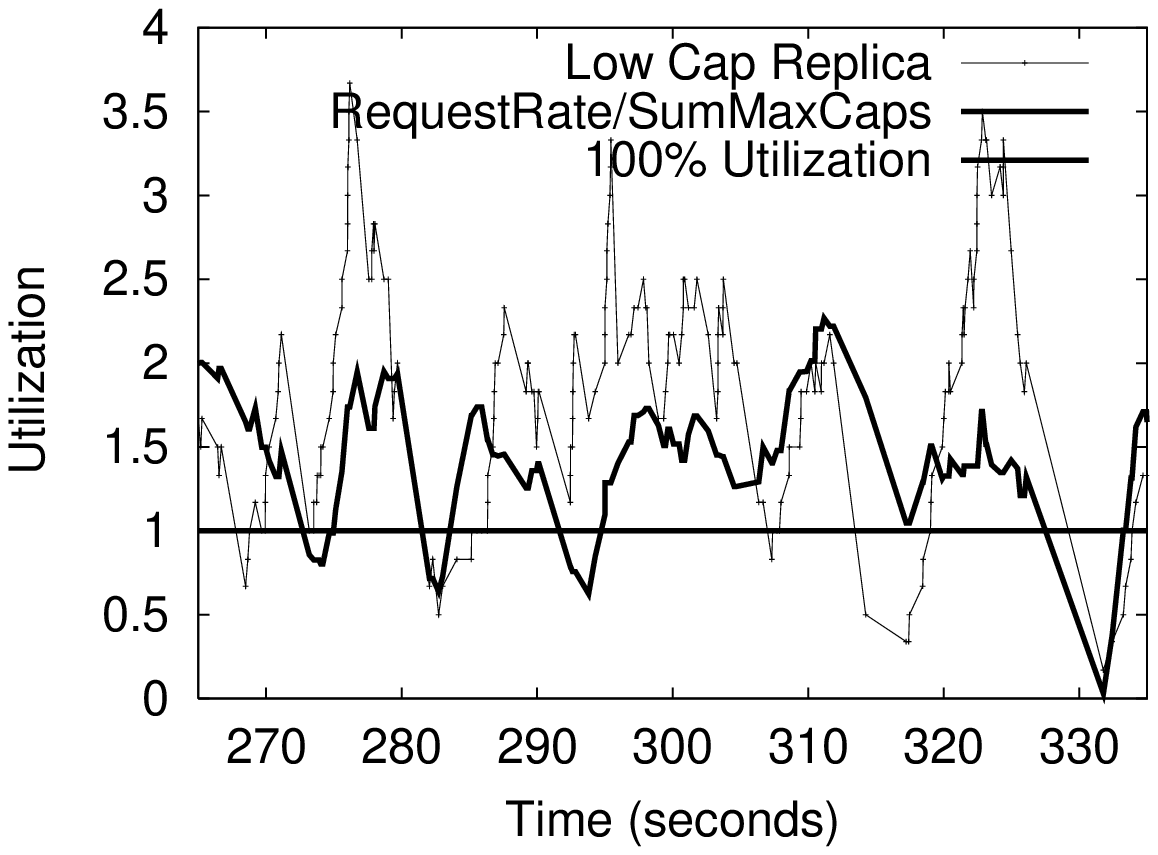}}
\caption{\small Low-capacity Replica Utilization versus Time for Max-Cap, Pareto arrivals.}
\label{Pareto-Repl1-Max-Cap}
\end{figure}

\begin{figure}
\centerline{\includegraphics[width=8cm]{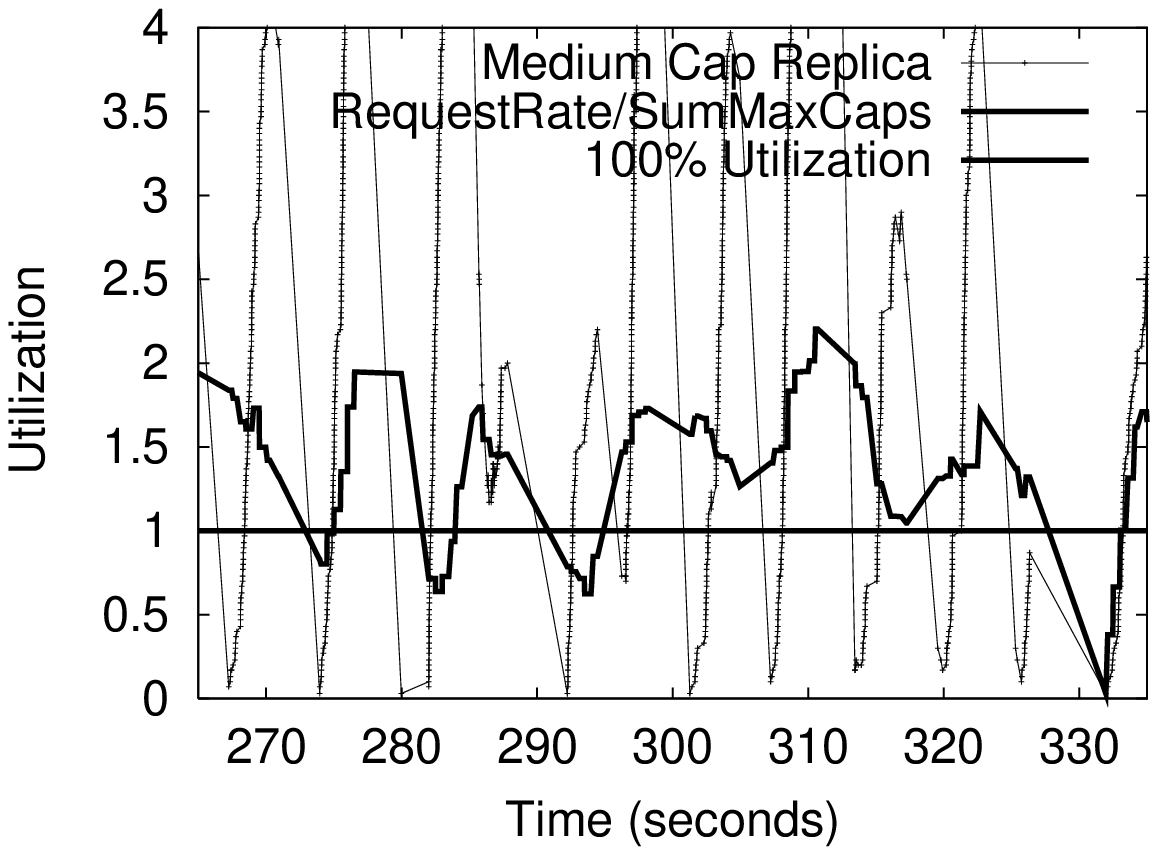}}
\caption{\small Medium-capacity Replica Utilization versus Time for Avail-Cap, Pareto arrivals.}
\label{Pareto-Repl3-Avail-Cap}
\end{figure}

\begin{figure}
\centerline{\includegraphics[width=8cm]{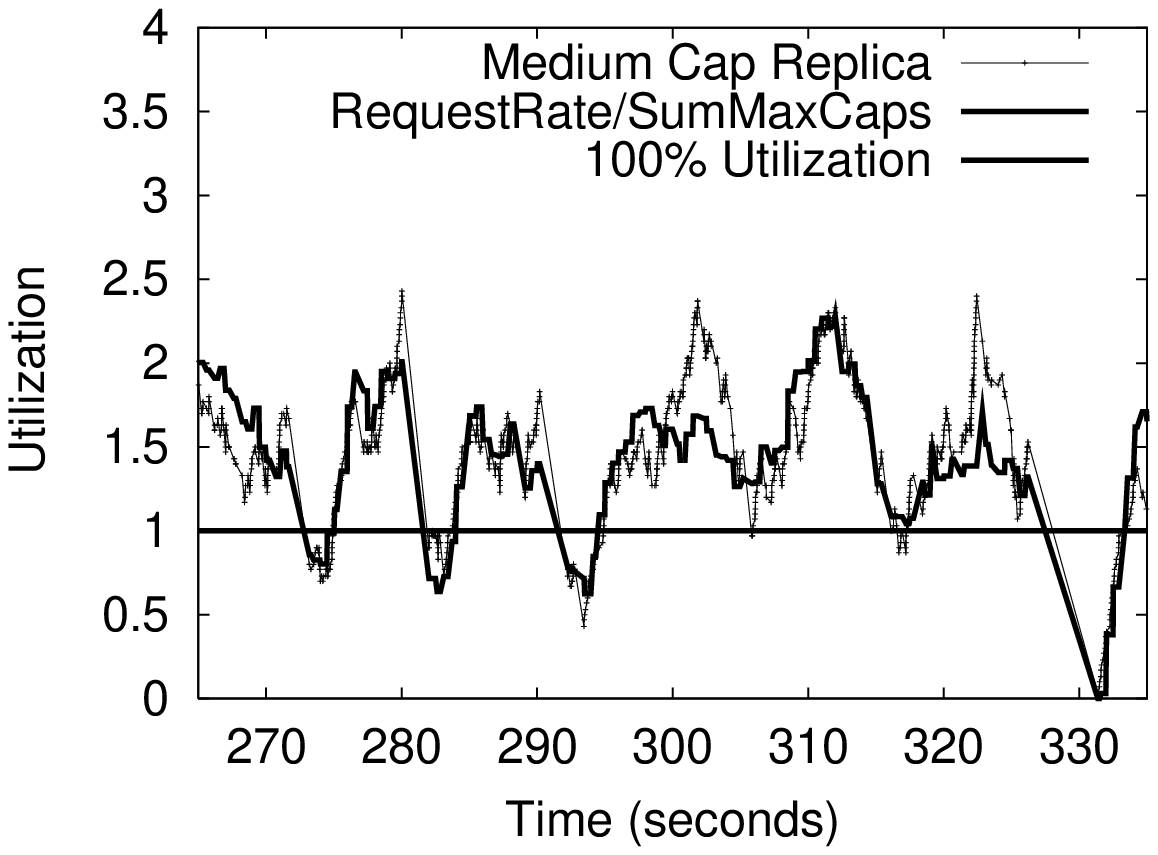}}
\caption{\small Medium-capacity Replica Utilization versus Time for Max-Cap, Pareto arrivals.}
\label{Pareto-Repl3-Max-Cap}
\end{figure}

\begin{figure}
\centerline{\includegraphics[width=8cm]{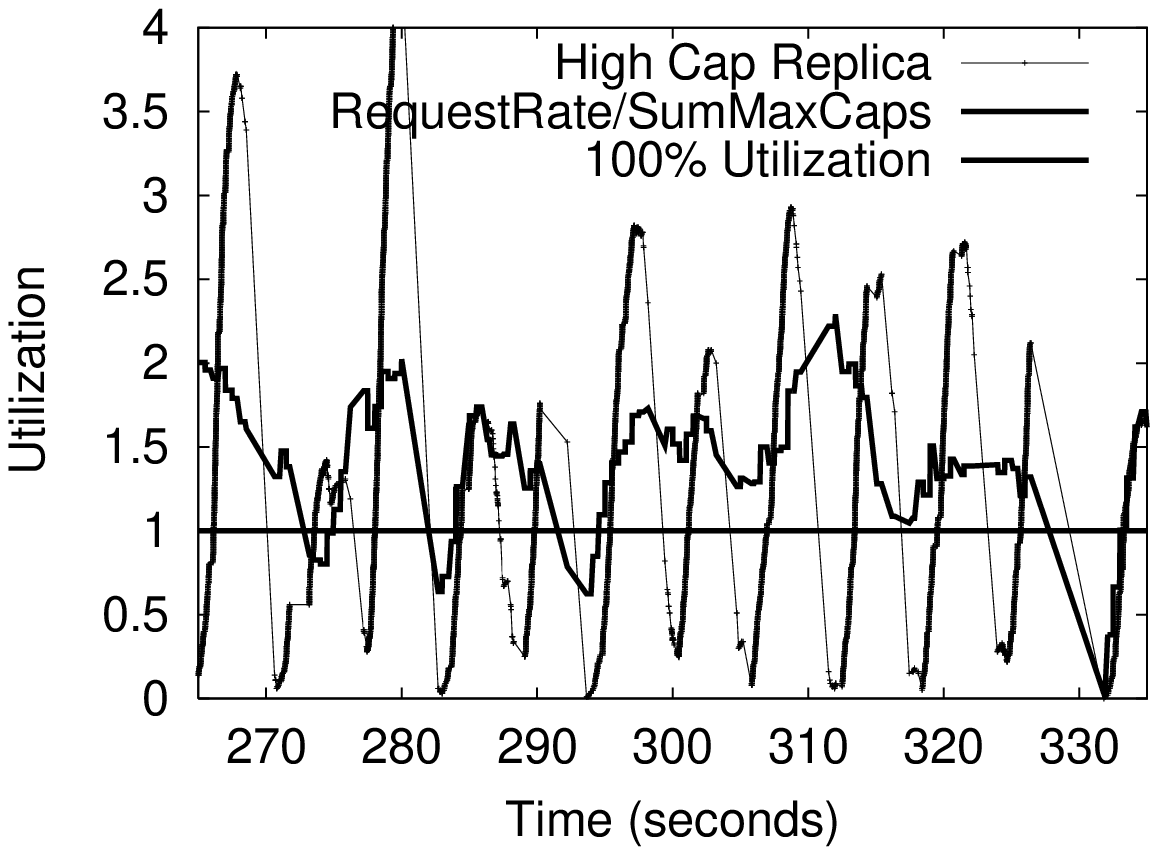}}
\caption{\small High-capacity Replica Utilization versus Time for Avail-Cap, Pareto arrivals.}
\label{Pareto-Repl9-Avail-Cap}
\end{figure}

\begin{figure}
\centerline{\includegraphics[width=8cm]{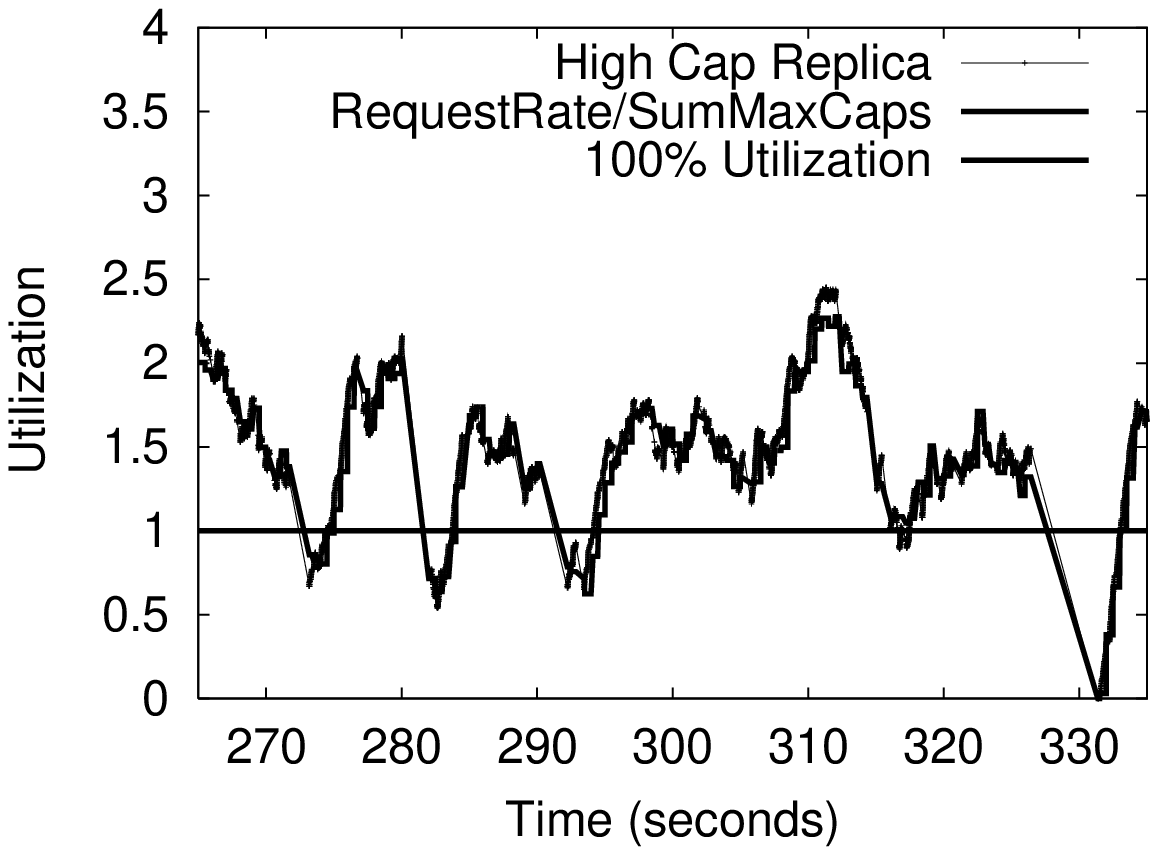}}
\caption{\small High-capacity Replica Utilization versus Time for Max-Cap, Pareto arrivals.}
\label{Pareto-Repl9-Max-Cap}
\end{figure}

In Figures~\ref{Pareto-Repl1-Avail-Cap}-\ref{Pareto-Repl9-Max-Cap} we
plot the same representative replica utilizations over a one minute
period in the experiment.  We also plot the ratio of the overall
request rate to the total maximum capacities as well as the $y=100\%$
utilization line.  From the figures we see that Avail-Cap suffers from
much wilder oscillation than Max-Cap, causing much higher peaks and
lower valleys in replica utilization than Max-Cap.  Moreover, Max-Cap
adjusts better to the fluctuations in the request rate; the
utilization curves for Max-Cap tend to follow the ratio curve more
closely than those for Avail-Cap.
 
(Note that idle periods contribute to the drops in utilization of
replicas in this experiment.  For example, an idle period occurs
between times 324 and 332 at which point we see a decrease in both the
ratio and the replica utilization.)

\subsubsection{Why Avail-Cap Can Suffer}

From the experiments above we see that Avail-Cap can suffer from
severe oscillation even when the overall request rate is well below
(e.g., 80\%) the total maximum capacities of the replicas.  The reason
why Avail-Cap does not balance load well here is that a vicious cycle
is created where the available capacity update of one replica affects
a subsequent update of another replica.  This in turn affects later
allocation decisions made by nodes which in turn affects later replica
updates.  This description becomes more concrete if we consider what
happens when a replica is overloaded.

In Avail-Cap, if a replica becomes overloaded, it reports an available
capacity of zero.
This report eventually reaches all peer nodes, causing them to stop
redirecting requests to the replica.  The exclusion of the overloaded
replica from the allocation decision shifts the entire burden of the
workload to the other replicas.  This can cause other replicas to
overload and report zero available capacity while the excluded replica
experiences a sharp decrease in its utilization.  This sharp decrease
causes the replica to begin reporting positive available capacity
which begins to attract requests again.  Since in the meantime other
replicas have become overloaded and excluded from the allocation
decision, the replica receives a flock of requests which cause it to
become overloaded again.  As we observed in previous sections, a
replica can experience wild and periodic oscillation where its
utilization continuously rises above its maximum capacity and falls
sharply.

In Max-Cap, if a replica becomes overloaded, the overload condition is
confined to that replica.
The same is true in the case of underloaded replicas.  Since the
overload/underload situations of the replicas are not reported, they
do not influence follow-up LBI updates of other replicas.  It is this
key property that allows Max-Cap to avoid herd behavior.

There are situations however where Avail-Cap performs well
without suffering from oscillation (see Section~\ref{DynamicRepSet}).
We next describe the factors that affect the performance of Avail-Cap
to get a clearer picture of when the reactive nature of  Avail-Cap is
beneficial (or at least not harmful) and when it causes oscillation.

\subsubsection{Factors Affecting Avail-Cap}
There are four factors that affect the performance of Avail-Cap: the
inter-update period $U$, the inter-request period $R$, the amount of
time $T$ it takes for all nodes in the network to receive the latest
update from a replica, and the ratio of the overall request rate to
the total maximum capacities of the replicas.  We examine these
factors by considering three cases:

\emph{Case 1:} $U$ is much smaller than $R$ ($U << R$), and $T$ is
sufficiently small so that when a replica pushes an update, all nodes
in the CUP tree receive the update before the next request arrival in
the network.  In this case, Avail-Cap performs well since all nodes
have the latest load-balancing information whenever they receive a
request.

\emph{Case 2:} $U$ is long relative to $R$ ($U > R$) and the overall
request rate is less than about 60\% the total maximum capacities of
the replicas.  (This 60\% threshold is specific to the particular
configuration of replicas we use: 10\% low, 60\% medium, 30\%
high. Other configurations have different threshold percentages that
are typically well below the total maximum capacities of the
replicas.)  In this case, when a particular replica overloads, the
remaining replicas are able to cover the proportion of requests
intended for the overloaded replica because there is a lot of extra
capacity in the system.  As a result, Avail-Cap avoids oscillations.
We see experimental evidence for this in
Section~\ref{DynamicRepSet}. However, over-provisioning to have enough
extra capacity in the system so that Avail-Cap can avoid oscillation
in this particular case seems a high price to pay for load stability.

\emph{Case 3: } $U$ is long relative to $R$ ($U > R$) and the overall
request rate is more than about 60\% the total maximum
capacities of the replicas.  In this case, as we observe in the
experiments above, Avail-Cap can suffer from oscillation.  This is
because every request that arrives directly affects the available
capacity of one of the replicas.  Since the request rate is greater
than the update rate, an update becomes stale shortly after a replica
has pushed it out.  However, the replica does not inform the nodes of
its changing available capacity until the end of its current update
period.  By that point many requests have arrived and have been
allocated using the previous, stale available capacity information.

In Case 3, Avail-Cap can suffer even if $T=0$ and updates were to
arrive at all nodes immediately after being issued.  This is because
all nodes would simultaneously exclude an overloaded replica from the
allocation decision until the next update is issued.  As $T$
increases, the staleness of the report only exacerbates the
performance of Avail-Cap.

In a large peer-to-peer network (more than 1000 nodes) we expect that
$T$ will be on the order of seconds since current peer-to-peer
networks with more than 1000 nodes have diameters ranging from a
handful to several hops \cite{ripeanu02}.  We consider $U$ = 1 second to be as
small (and aggressive) an inter-update period as is practical in a
peer-to-peer network.  In fact even one second may be too aggressive
due to the overhead it generates.  This means that when particular
content experiences high popularity, we expect that typically $U+T >>
R$.  Under such circumstances Avail-Cap is not a good load-balancing
choice.  For less popular content, where $U+T < R$, Avail-Cap is a
feasible choice, although it is unclear whether load-balancing across
the replicas is as urgent here, since the request rate is low.

The performance of Max-Cap is independent of the values of $U$, $R$,
and $T$.  More importantly, Max-Cap does not require continuous
updates; replicas issue updates only if they choose to re-issue new
contracts to report changes in their maximum capacities.  (See Section
~\ref{ExtraneousLoad}).  Therefore, we believe that Max-Cap is a more
practical choice in a peer-to-peer context than Avail-Cap.

\subsection{Dynamic Replica Set}
\label{DynamicRepSet}

A key characteristic of peer-to-peer networks is that they are subject
to constant change; peer nodes continuously enter and leave the
system.  In this experiment we compare Max-Cap with Avail-Cap
when replicas enter and leave the system.  We present
results here for a Poisson request arrival rate that is 80\% the
total maximum capacities of the replicas.

We present two dynamic experiments.  In both experiments, the network
starts with ten replicas and after a period of 600 seconds, movement
into and out of the network begins.  In the first experiment, one replica
leaves and one replica enters the network every 60 seconds.  In the
second and much more dynamic experiment, five replicas leave and five
replicas enter the network every 60 time units.  The replicas that
leave are randomly chosen.  The replicas that enter the network enter
with maximum capacities of 1, 10, and 100 with probability of 0.10, 0.60,
and 0.30 respectively as in the initial allocation.  This means that
the total maximum capacities of the active replicas in the network
varies throughout the experiment, depending on the capacities of the
entering replicas. 

\begin{figure}
\centerline{\includegraphics[width=8cm]{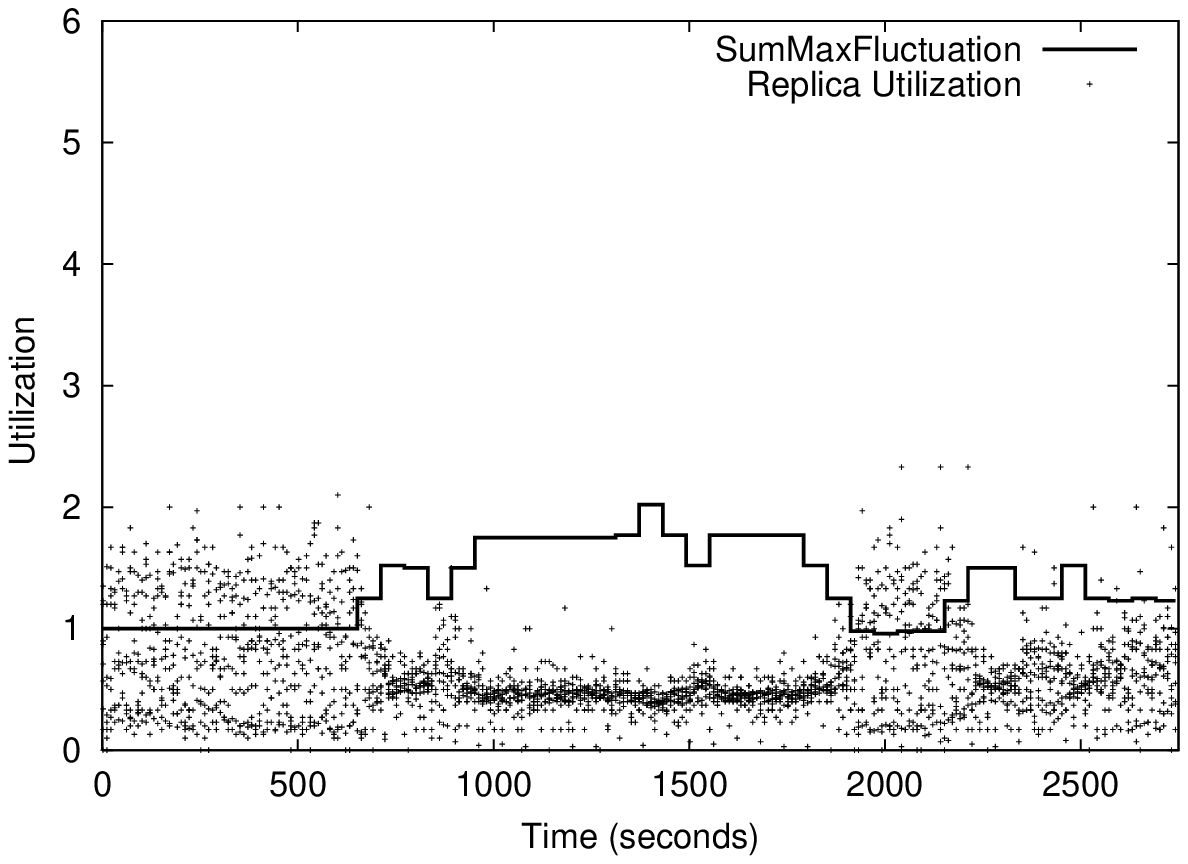}}
\caption{\small Replica Utilization versus Time for Avail-Cap with a dynamic replica set.  One replica enters and leaves every 60 seconds.}
\label{Dynamic-Scatterplot-10-1-60-Avail-Cap}
\end{figure}

\begin{figure}
\centerline{\includegraphics[width=8cm]{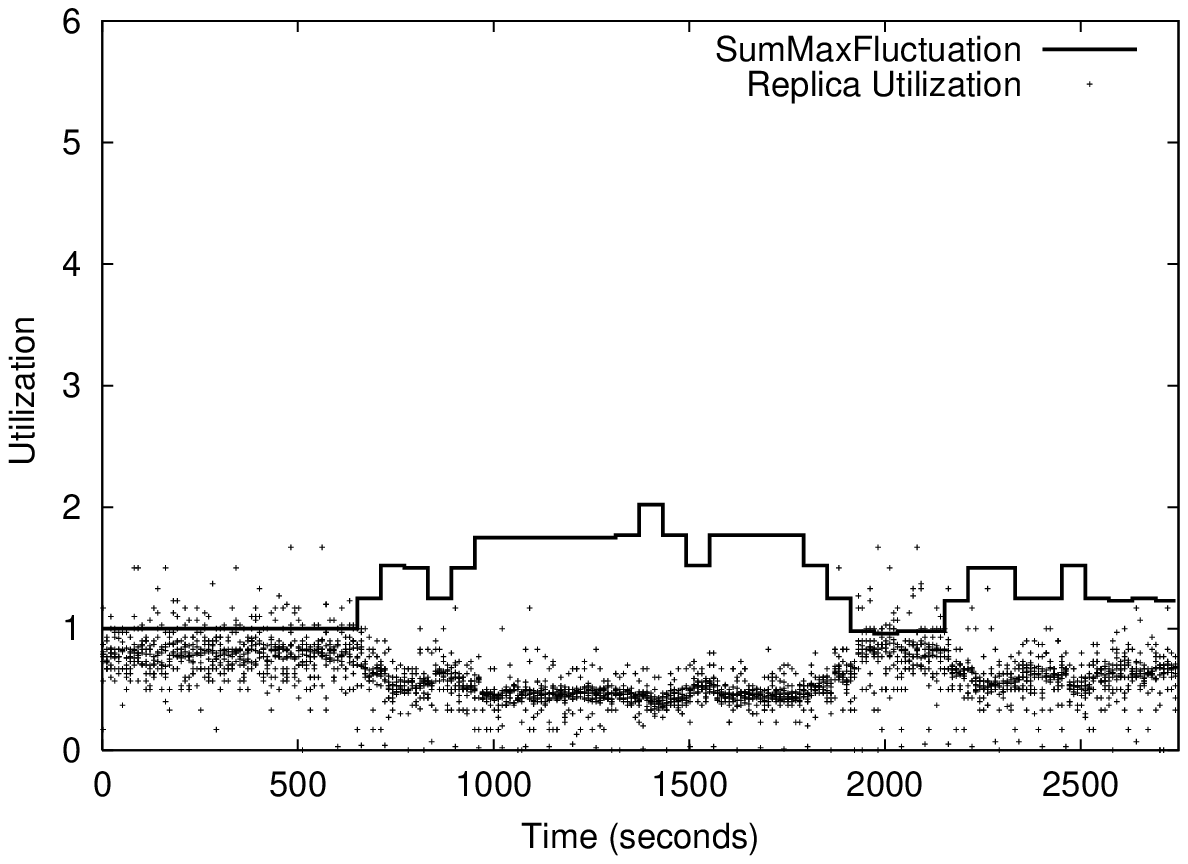}}
\caption{\small Replica Utilization versus Time for Max-Cap with a dynamic replica set.  One replica enters and leaves every 60 seconds.}
\label{Dynamic-Scatterplot-10-1-60-Max-Cap}
\end{figure}

Figures~\ref{Dynamic-Scatterplot-10-1-60-Avail-Cap} and
\ref{Dynamic-Scatterplot-10-1-60-Max-Cap} show for the first dynamic
experiment the utilization of active replicas throughout time as
observed for Avail-Cap and Max-Cap.  Note that points with zero
utilization indicate newly entering replicas.
The jagged line plots the ratio of the current sum of maximum
capacities in the network, $S_{curr}$, to the original sum of maximum
capacities, $S_{orig}$.  With each change in the replica set, the
replica utilizations for both Avail-Cap and Max-Cap change.  Replica
utilizations rise when $S_{curr}$ falls and vice versa.

From the figure we see that between times 1000 and 1820, $S_{curr}$ is
between 1.75 and 2 times $S_{orig}$, and is more than double the
overall workload rate of $0.8 * S_{orig}$.  During this time period,
Avail-Cap performs quite well because the workload rate is not very
demanding and there is plenty of extra capacity in the system (Case 2
above).  However, when at time 1940 $S_{curr}$ falls back to
$S_{orig}$, we see that both algorithms exhibit the same behavior as
they do at the start, between times 0 and 600.  Max-Cap readjusts
nicely and clusters replica utilization at around 80\%, while
Avail-Cap starts to suffer again.

\begin{figure}
\centerline{\includegraphics[width=8cm]{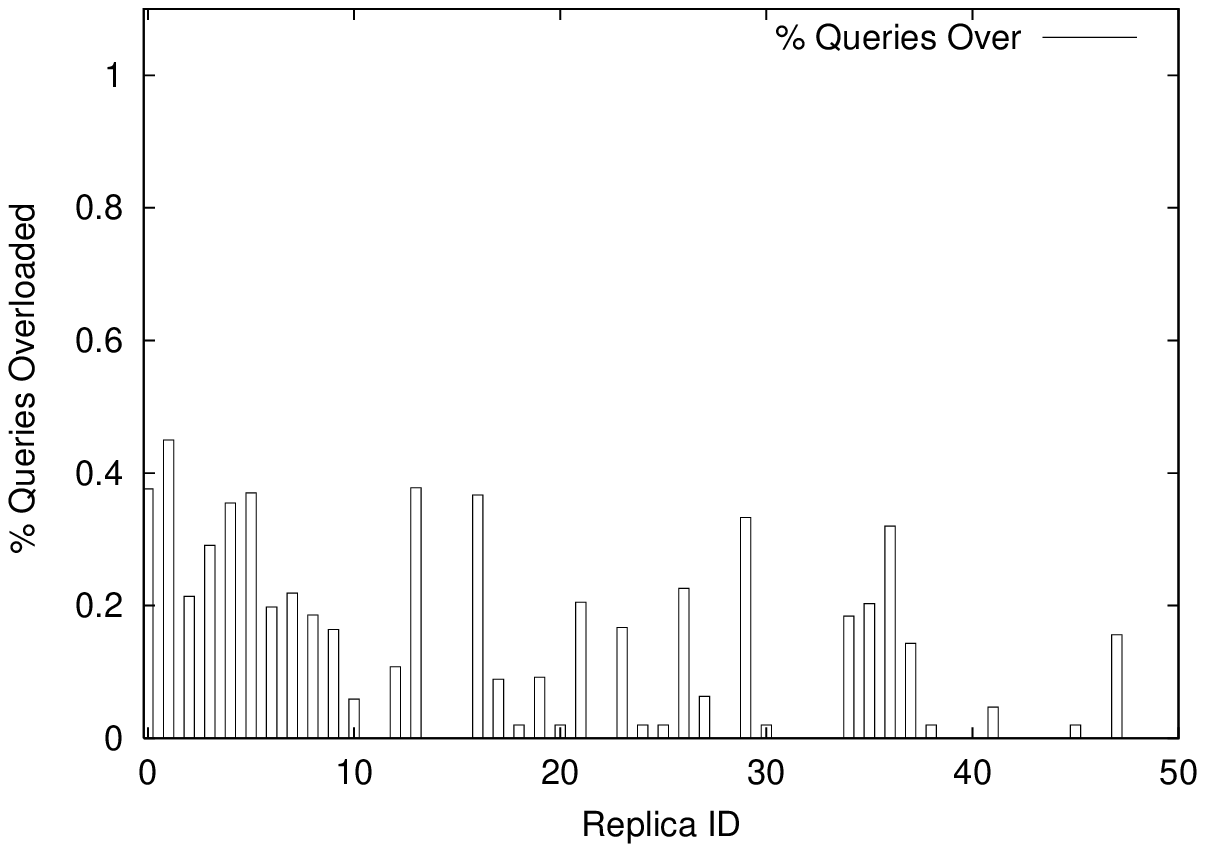}}
\caption{\small Percentage Overloaded Queries versus Replica ID for Avail-Cap with a dynamic replica set.  One replica enters and leaves every 60 seconds.}
\label{Dynamic-Overloaded-10-1-60-Avail-Cap}
\end{figure}

\begin{figure}
\centerline{\includegraphics[width=8cm]{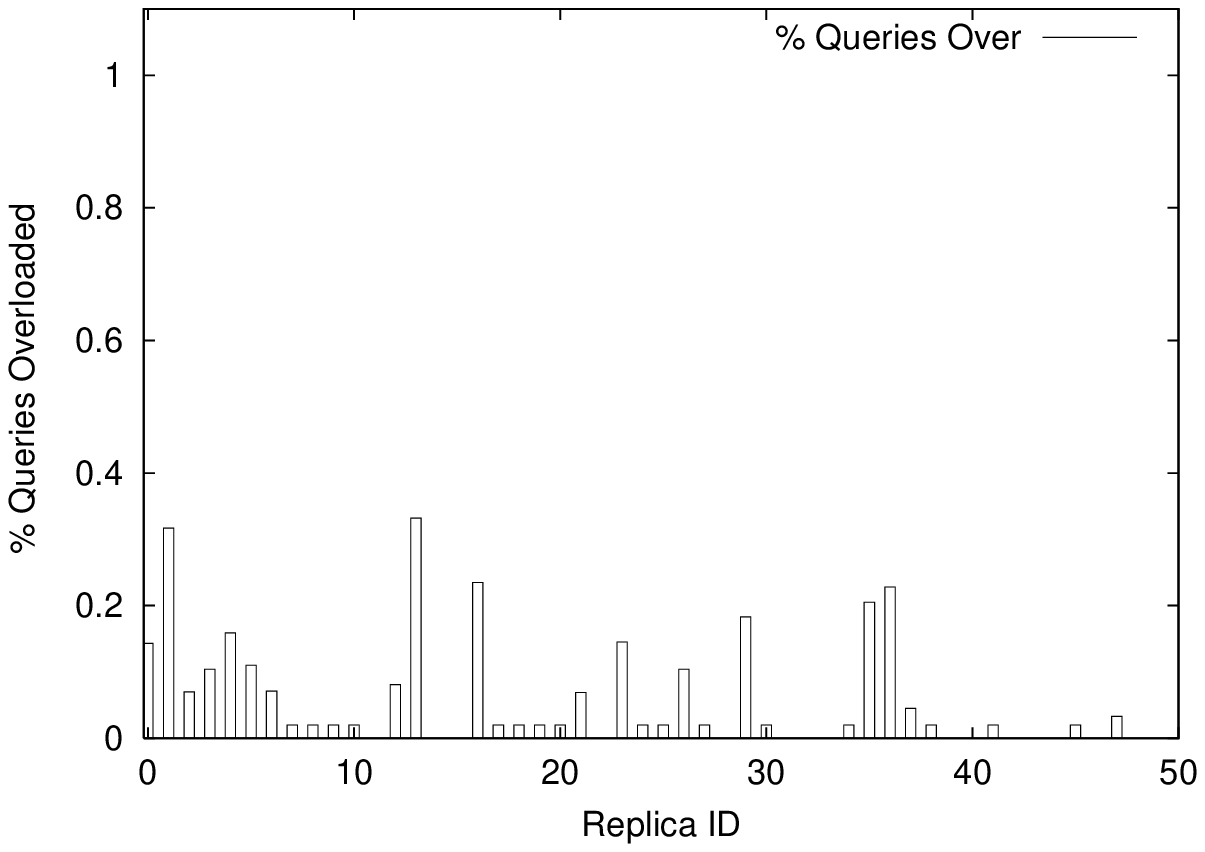}}
\caption{\small Percentage Overloaded Queries versus Replica ID for Max-Cap with a dynamic replica set.  One replica enters and leaves every 60 seconds.}
\label{Dynamic-Overloaded-10-1-60-Max-Cap}
\end{figure}

Figures~\ref{Dynamic-Overloaded-10-1-60-Avail-Cap} and
\ref{Dynamic-Overloaded-10-1-60-Max-Cap} show for the first dynamic
experiment the percentage of queries that were received by each replica
while the replica was overloaded for Avail-Cap and Max-Cap.  Replicas
that entered and departed the network throughout the simulation were
chosen from a pool of 50 replicas.  Those replicas in the pool which
did not participate in this experiment do not have a bar associated
with their ID in the figure.  From the figure, we see that Max-Cap
achieves smaller overload query percentages across all replica IDs.

\begin{figure}[t]
\centerline{\includegraphics[width=8cm]{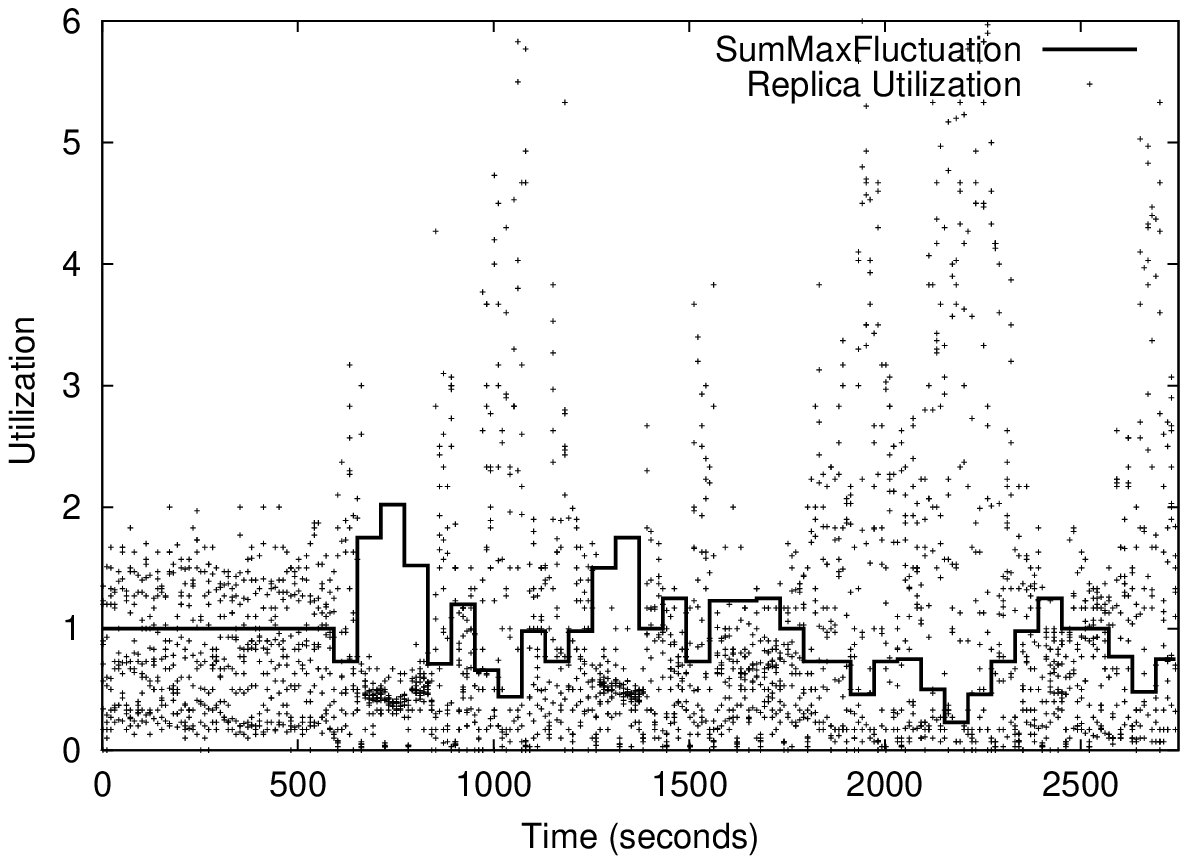}}
\caption{\small Replica Utilization versus Time for Avail-Cap with a dynamic replica set.  Half the replicas enter and leave every 60 seconds.}
\label{Dynamic-Scatterplot-10-5-60-Avail-Cap}
\end{figure}

\begin{figure}
\centerline{\includegraphics[width=8cm]{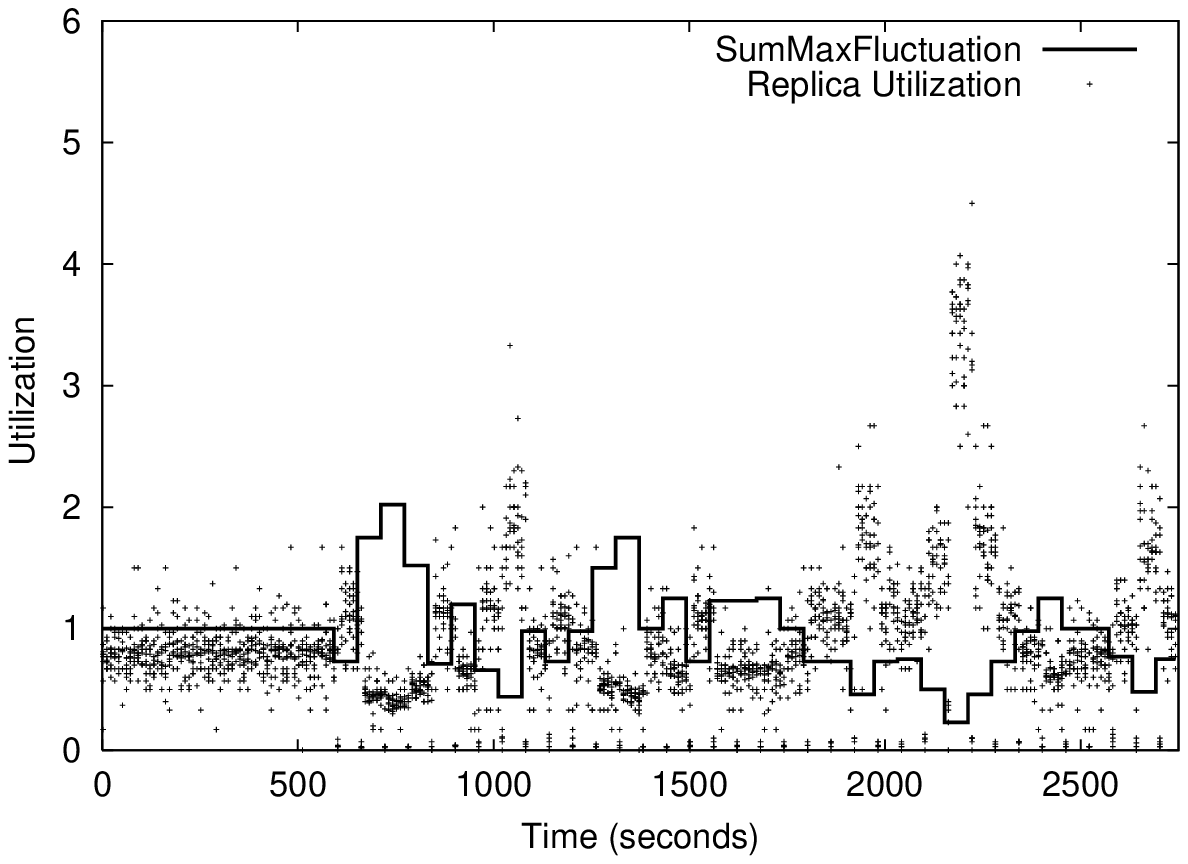}}
\caption{\small Replica Utilization versus Time for Max-Cap with a dynamic replica set.  Half the replicas enter and leave every 60 seconds.}
\label{Dynamic-Scatterplot-10-5-60-Max-Cap}
\end{figure}

\begin{figure}
\centerline{\includegraphics[width=8cm]{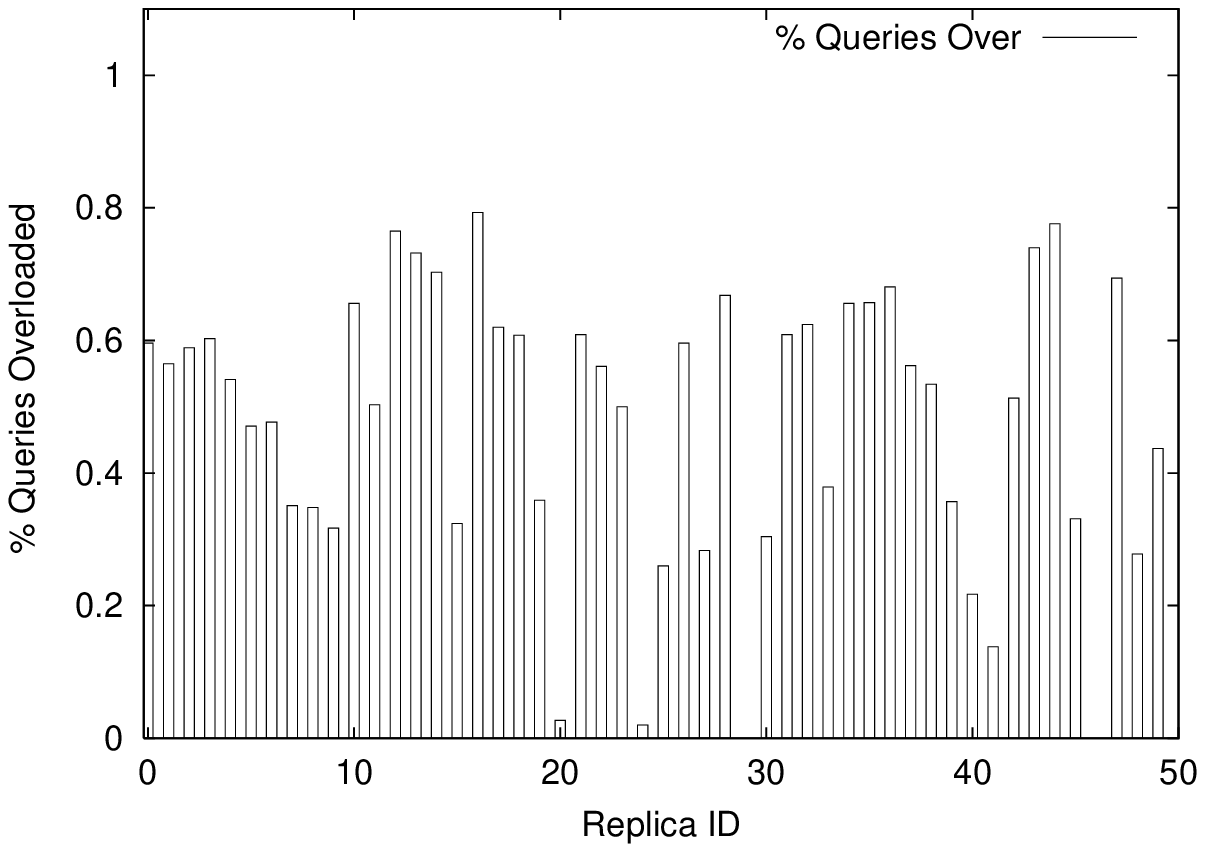}}
\caption{\small Percentage Overloaded Queries versus Replica ID for Avail-Cap with a dynamic replica set.  Half the replicas enter and leave every 60 seconds.}
\label{Dynamic-Overloaded-10-5-60-Avail-Cap}
\end{figure}

\begin{figure}
\centerline{\includegraphics[width=8cm]{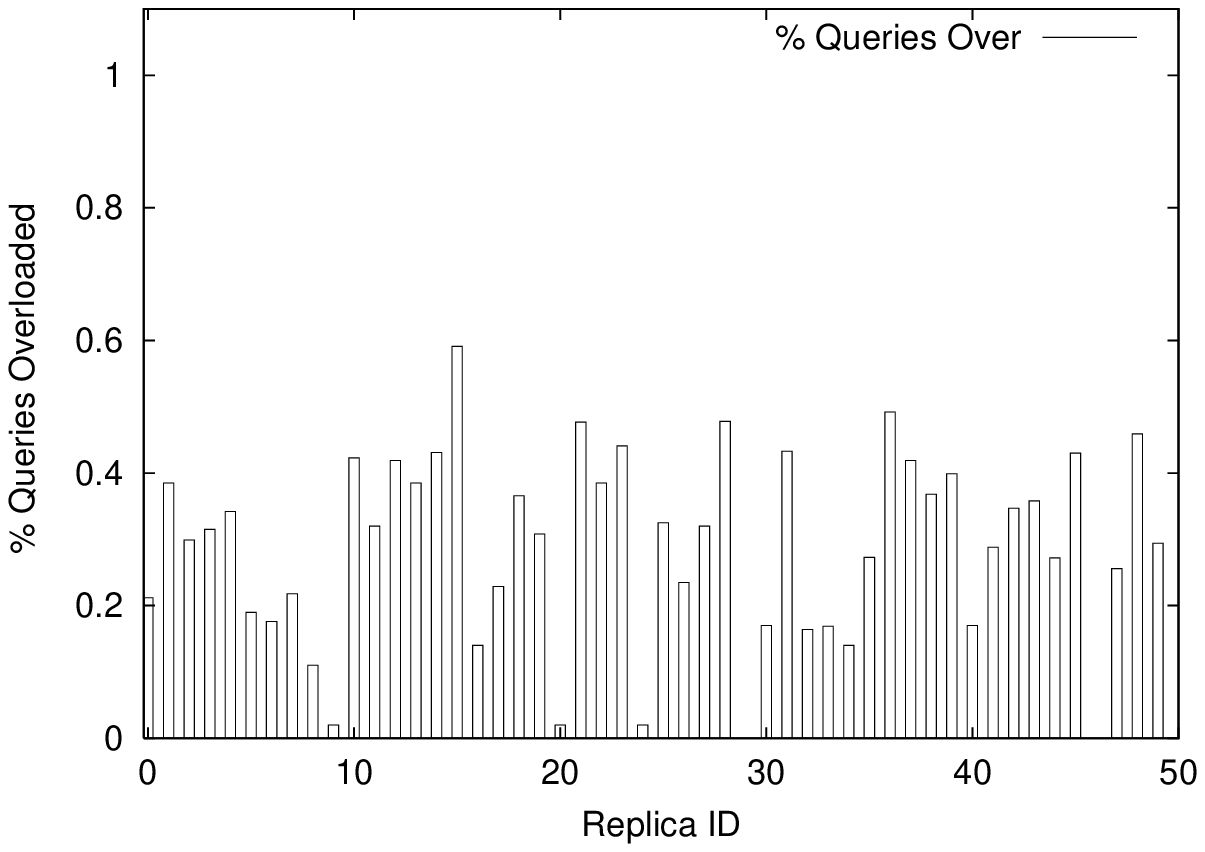}}
\caption{\small Percentage Overloaded Queries versus Replica ID for Max-Cap with a dynamic replica set.  Half the replicas enter and leave every 60 seconds. }
\label{Dynamic-Overloaded-10-5-60-Max-Cap}
\end{figure}

Figures~\ref{Dynamic-Scatterplot-10-5-60-Avail-Cap} and
\ref{Dynamic-Scatterplot-10-5-60-Max-Cap} show the utilization
scatterplot and Figures~\ref{Dynamic-Overloaded-10-5-60-Avail-Cap} and
\ref{Dynamic-Overloaded-10-5-60-Max-Cap} show the overloaded query
percentage for the second dynamic experiment.  We see that changing
half the replicas every 60 seconds can dramatically affect $S_{curr}$.
For example, when $S_{curr}$ drops to $0.2 S_{orig}$ at time 2161, we
see the utilizations rise dramatically for both Avail-Cap and Max-Cap.
This is because during this period the workload rate is four times
that of $S_{curr}$.  However by time 2401, $S_{curr}$ has risen to
$1.2 S_{orig}$ which allows for both Avail-Cap and Max-Cap to adjust
and decrease the replica utilization.  At the next replica set change
at time 2461, $S_{curr}$ equals $S_{orig}$.  During the next minute we
see that Max-Cap overloads very few replicas whereas Avail-Cap does
not recuperate as well.  Similarly, when examining the overloaded
query percentage we see that Max-Cap achieves smaller percentages when
compared with Avail-Cap.

The two dynamic experiments we have described above show two things;
first, when the workload is not very demanding and there is unused
capacity, the behaviors of Avail-Cap and Max-Cap are similar
However, Avail-Cap suffers more as overall available capacity
decreases.  Second, Avail-Cap is affected more by short-lived
fluctuations (in particular, decreases) in total maximum capacity than
Max-Cap.  This is because the reactive nature of Avail-Cap causes it
to adapt abruptly to changes in capacities, even when these changes
are short-lived.

\subsection{Extraneous Load}
\label{ExtraneousLoad}

When replicas can honor their maximum capacities, Max-Cap avoids the
oscillation that Avail-Cap can suffer, and does so with no update
overhead.  Occasionally, some replicas may not be able to honor their
maximum capacities because of \emph{extraneous load} caused by other
applications running on the replicas or network conditions unrelated
to the content request workload.

To deal with the possibility of extraneous load, we modify the Max-Cap
algorithm slightly to work with honored maximum capacities.  A
replica's honored maximum capacity is its maximum capacity minus the
extraneous load it is experiencing.  The algorithm changes slightly; a
peer node chooses a replica to which to forward a content request with
probability proportional to the honored maximum capacity advertised by
the replica.  This means that replicas may choose to send updates to
indicate changes in their honored maximum capacities.  However, as we
will show, the behavior of Max-Cap is not tied to the timeliness of
updates in the way Avail-Cap is.

We view the honored maximum capacity reported by a replica as a
contract.  If the replica cannot adhere to the contract or has extra
capacity to give, but does not report the deficit or surplus, then
that replica alone will be affected and may be overloaded or
underloaded since it will be receiving a request share that is
proportional to its previous advertised honored maximum capacity.

If, on the other hand, a replica chooses to issue a new contract with
the new honored maximum capacity, then this new update can affect the
load balancing decisions of the nodes in the peer network and the
workload could shift to the other replicas. This shift in workload is
quite different from that experienced by Avail-Cap when a replica
reports overload and is excluded.  The contracts of any other replica
will not be affected by this workload shift.  Instead, the contract is
are solely affected by the extraneous load that replica experiences
which is independent of the extraneous load experienced by the replica
issuing the new contract.  This is unlike Avail-Cap where the
available capacity reported by one replica directly affects the
available capacities of the others.

\begin{figure}
\centerline{\includegraphics[width=8cm]{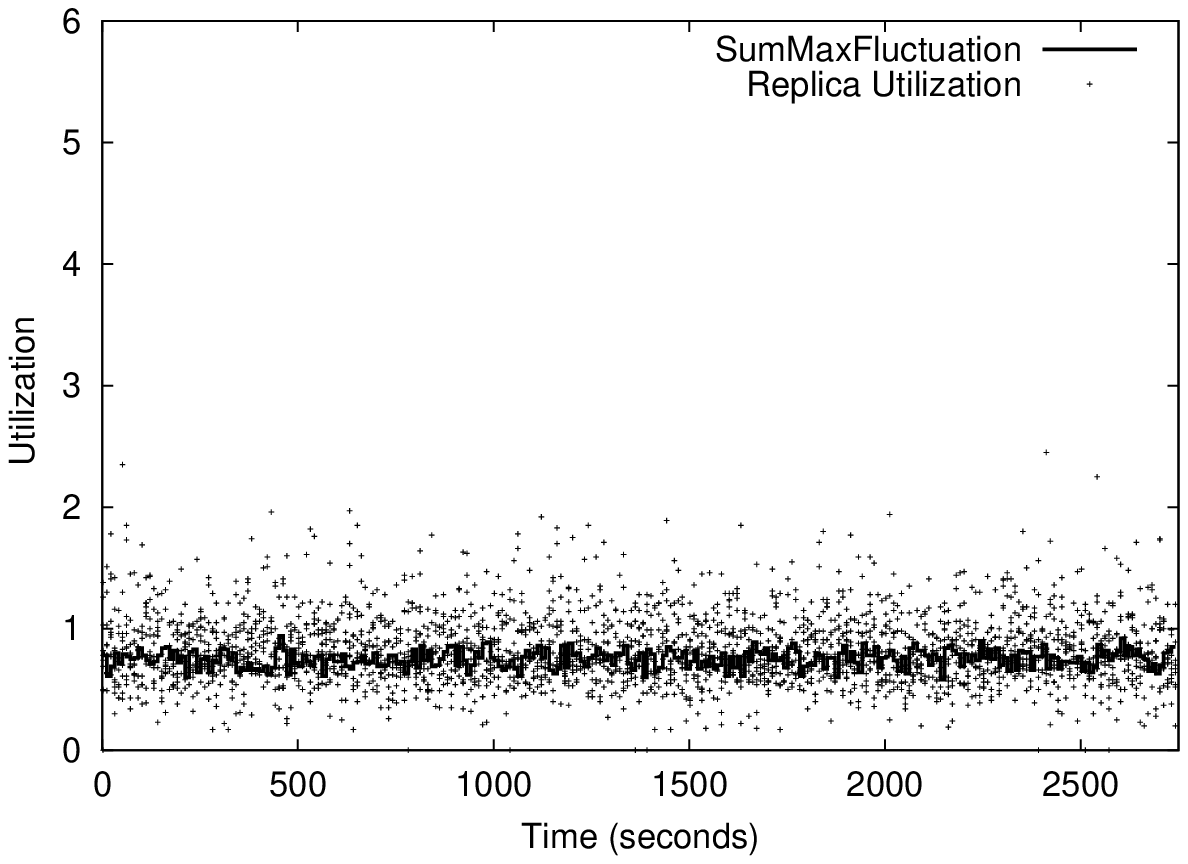}}
\caption{\small Replication Utilization versus Time for Max-Cap with extraneous load and an inter-update period of one second.}
\label{Xload-Max-Cap-pd-1-upTo50}
\end{figure}

\begin{figure}
\centerline{\includegraphics[width=8cm]{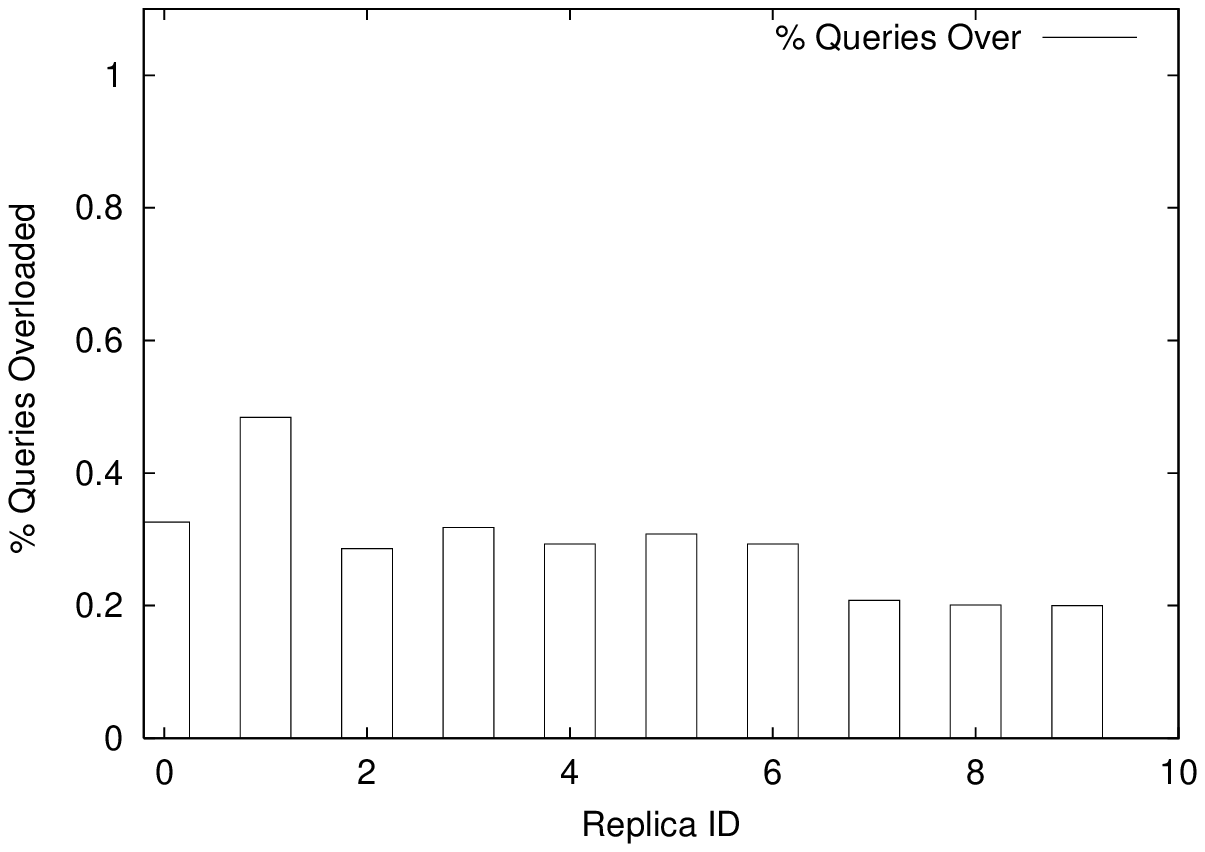}}
\caption{\small Percentage Overloaded Queries versus Replica ID, Max-Cap with extraneous load and an inter-update period of one second.}
\label{Xload-OverCapQu-Max-Cap-pd-1-upTo50}
\end{figure}

In this section we study the performance of Max-Cap in an experiment
where all replica nodes are continuously issuing new
contracts. Specifically, for each of ten replicas, we inject
extraneous load into the replica once a second.  The extraneous load
injected is randomly chosen to be anywhere between 0\% and 50\% of
the replica's original maximum capacity.
Figures~\ref{Xload-Max-Cap-pd-1-upTo50} and
\ref{Xload-OverCapQu-Max-Cap-pd-1-upTo50} show the replica utilization
versus time and the overloaded query percentages for Max-Cap with an
inter-update period of 1 second.  The jagged line in
Figure~\ref{Xload-Max-Cap-pd-1-upTo50} shows the total honored maximum
capacities over time.  Since throughout the experiment each replica's
honored maximum capacity varies between 50\% and 100\% its
original maximum capacity, the total maximum capacity is expected to
hover at around 75\% the original total maximum capacity and we
see that the jagged line hovers around this value.  We therefore
generate Poisson request arrivals with an average rate that is 80\%
of this value to keep consistent with our running example of
80\% workload rates.

From the figures, we see that Max-Cap continues to cluster replica
utilization at around 80\%, but there are more overloaded replicas
throughout time than when compared with the experiment in which all
replicas adhere to their contracts all the time
(Figure~\ref{Max-Cap-289-Scatterplot}).  We also see that the
overloaded percentages are higher than before
(Figure~\ref{Max-Cap-289-OverCapQueries}).  The reason for this
performance degradation is that the randomly injected load (of 0\% to
50\%) can cause sharp rises and falls in the reported contract of each
replica from one second to the next.  Since the change is so rapid,
and updates take on the order of seconds to reach all allocating
nodes, allocation decisions are continuously being made using stale
information.

\begin{figure}
\centerline{\includegraphics[width=8cm]{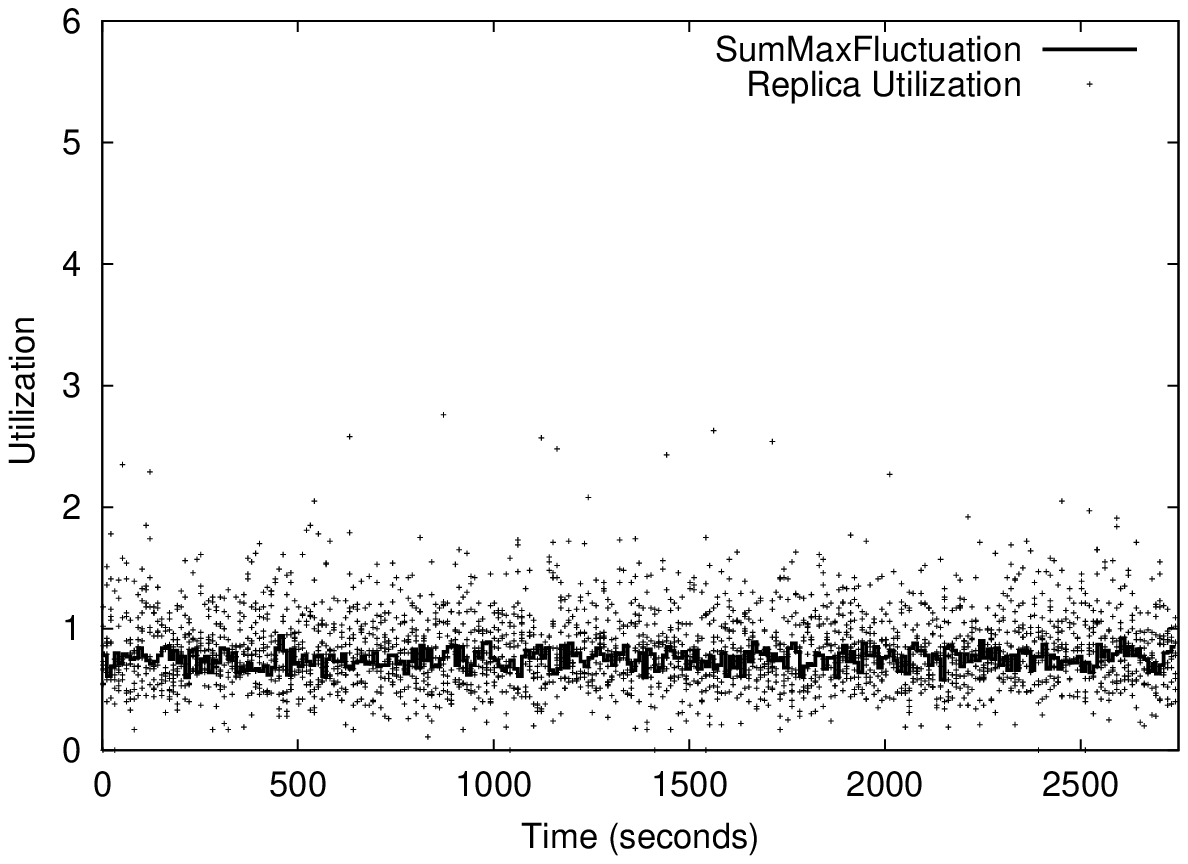}}
\caption{\small Replication Utilization versus Time for Max-Cap with extraneous load and an inter-update period of ten seconds.}
\label{Xload-Max-Cap-pd-10-upTo50}
\end{figure}

\begin{figure}
\centerline{\includegraphics[width=8cm]{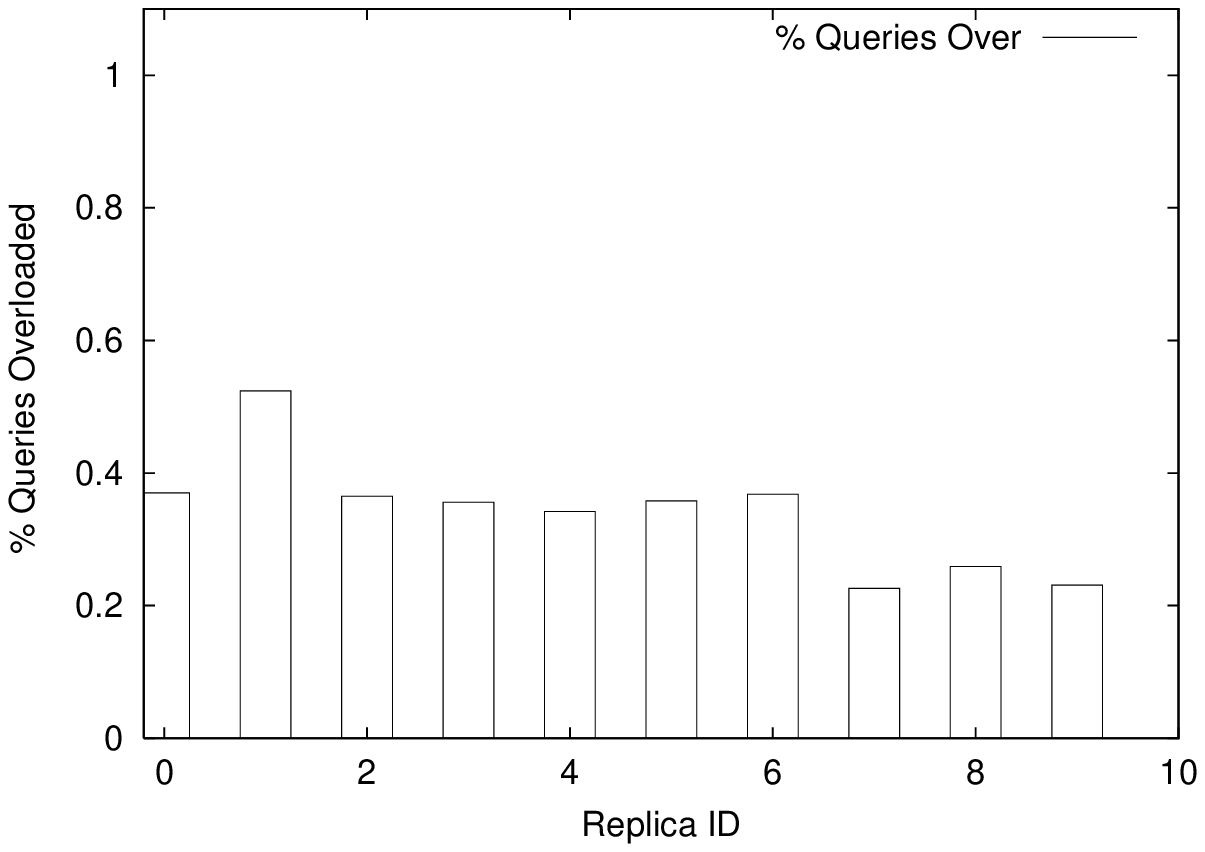}}
\caption{\small Percentage Overloaded Queries versus Replica ID for Max-Cap with extraneous load and an inter-update period of ten seconds.}
\label{Xload-OverCapQu-Max-Cap-pd-10-upTo50}
\end{figure}

In the next experiment we use the same parameters as above but we
change the update period to 10 seconds.
Figures~\ref{Xload-Max-Cap-pd-10-upTo50} and
\ref{Xload-OverCapQu-Max-Cap-pd-10-upTo50} show the utilization and
overloaded percentages for this experiment.  We see that the
overloaded percentages increase only slightly while the overhead of
pushing the updates decreases by a factor of ten.  In contrast, when
we perform the same experiment for Avail-Cap, we find that the
overloaded query percentages for Avail-Cap increase from about 55 to
more than 80\% across all the replicas when the inter-update period
changes from 1 to 10 seconds.  However, this performance degradation
is not so much due to the fluctuation of the extraneous load as it is
due to Avail-Cap's tendency to oscillate when the request rate is
greater than the update rate.

We purposely choose this scenario to test how Max-Cap performs under
widely fluctuating extraneous load on every replica.  We generally
expect that extraneous load will not fluctuate so wildly, nor will all
replicas issue new contracts every second.  Moreover, we expect
the inter-update period to be on the order of several seconds or even
minutes, which further reduces overhead. 

We can view the effect of extraneous load on the performance of
Max-Cap as similar to that seen in the dynamic replica experiments.
When a replica advertises a new honored maximum capacity, it is as if
that replica were leaving and being replaced by a new replica with a
different maximum capacity.

\section{Related Work}
\label{LBRelatedWork}

Load-balancing has been the focus of many studies described in the
distributed systems literature.  We first describe load-balancing
techniques that could be applied in a peer-to-peer context.  We
classify these into two categories, those algorithms where the
allocation decision is based on load and those where the allocation
decision is based on available capacity.  We then describe other
load-balancing techniques (such as process migration) that cannot be
directly applied in a peer-to-peer context.

\subsection{Load-Based Algorithms}
Of the load-balancing algorithms based on load, a very common approach
to performing load-balancing is to choose the server with the least
reported load from among a set of servers.  This approach performs
well in a homogeneous system where the task allocation is performed by
a single centralized entity (dispatcher) which has complete up-to-date
load information \cite{weber78,winston77}.  In a system where multiple
dispatchers are independently performing the allocation of tasks, this
approach however has been shown to behave badly, especially if load
information used is stale
\cite{eager86,mirchandaney89,mitzenmacher97,shivaratri92}.
Mitzenmacher talks about the ``herd behavior'' that can occur when
servers that have reported low load are inundated with requests from
dispatchers until new load information is reported
\cite{mitzenmacher97}.

Dahlin proposes \emph{load interpretation} algorithms \cite{dahlin99}.
These algorithms take into account the age (staleness) of the load
information reported by each of a set of distributed homogeneous
servers as well as an estimate of the rate at which new requests
arrive at the whole system to determine to which server to allocate a
request.

Many studies have focused on the strategy of using a subset of the
load information available.  This involves first randomly choosing a
small number, $k$, of homogeneous servers and then choosing the least
loaded server from within that set
\cite{mitzenmacher96,eager86,vved96,azar94,karp92}.  In particular,
for homogeneous systems, Mitzenmacher \cite{mitzenmacher96} studies the
tradeoffs of various choices of $k$ and various degrees of staleness
of load information reported.  As the degree of staleness increases,
smaller values of $k$ are preferable.

Genova et al. \cite{genova00} propose an algorithm, which we call
\emph{Inv-Load} that first randomly selects $k$ servers.  The
algorithm then weighs the servers by load information and chooses a
server with probability that is inversely proportional to the load
reported by that server.  When $k=n$, where $n$ is the total number of
servers, the algorithm is shown to perform better than previous
load-based algorithms and for this reason we focus on this algorithm
in this paper.

As we see in Section~\ref{Inv-Load-Heterogeneity}, algorithms that
base the decision on load do not handle heterogeneity.

\subsection{Available-Capacity-Based Algorithms}
Of the load-balancing algorithms based on available capacity, one
common approach has been to choose amongst a set of servers based on
the available capacity of each server \cite{zhu98} or the available
bandwidth in the network to each server \cite{carter97}.  The server
with the highest available capacity/bandwidth is chosen by a client
with a request. The assumption here is that the reported available
capacity/bandwidth will continue to be valid until the chosen server
has finished servicing the client's request.  This assumption does not
always hold; external traffic caused by other applications can
invalidate the assumption, but more surprisingly the traffic caused by
the application whose workload is being balanced can also invalidate
the assumption.  We see this in Section~\ref{Oscillation}.

Another approach is to to exclude servers that fail some utilization
threshold and to choose from the remaining servers.  Mirchandaney et
al. \cite{mirchandaney90} and Shivaratri et al. \cite{shivaratri92}
classify machines as lightly-utilized or heavily-utilized and then
choose randomly from the lightly-utilized servers.  This work focuses
on local-area distributed systems.  Colajanni et al. use this approach
to enhance round-robin DNS load-balancing across a set of widely
distributed heterogeneous web servers \cite{colajanni98},
Specifically, when a web server surpasses a utilization threshold it
sends an alarm signal to the DNS system indicating it is out of
commission.  The server is excluded from the DNS resolution until it
sends another signal indicating it is below threshold and free to
service requests again.  In this work, the maximum capacities of the
most capable servers are at most a factor of three that of the least
capable servers.

As we see in Section~\ref{Oscillation}, when applied in the context of
a peer-to-peer network where many nodes are making the allocation
decision and where the maximum capacities of the replica nodes can
differ by two orders of magnitude, excluding a serving node
temporarily from the allocation decision can result in load
oscillation.

\subsection{Other Load-balancing Techniques}
We now describe load-balancing techniques that appear in the literature but
cannot be directly applied in a peer-to-peer context. 

There has been a large body of work devoted to the problem of
load-balancing across a set of servers residing within a cluster.  In
some cluster systems there is one centralized dispatcher through which
all incoming requests to the cluster arrive.  The dispatcher has full
control over the allocation of requests to servers
\cite{dias96,cisco97}. In other systems there are multiple dispatchers
that make the allocation decision.  One common approach is to have
front-end servers sit at the entrance of the cluster intercepting
incoming requests and allocating requests to the back-end servers
within the cluster that actually satisfy the requests
\cite{castro99}. Still others have requests be evenly routed to
servers within the cluster via DNS rotation (described below) or via a
single IP-switch sitting at the front of the cluster (e.g.,
\cite{foundry98}).  Upon receiving a request each server then decides
whether to satisfy the request or to dispatch it to another server
\cite{zhu98}.  Some cluster systems have the dispatchers(s) poll each
server or a random set of servers for load/availability information
just before each allocation decision \cite{awerbuch96,shen02}.  Others
have the dispatcher(s) periodically poll servers, while still others
have servers periodically broadcast their load-balancing information.
Studies that compare the tradeoffs among these information
dissemination options within a cluster include \cite{zhu98,shen02}.

Regardless of the way this information is exchanged, cluster-based
algorithms take advantage of the local-area nature of the cluster
network to deliver timely load-balancing updates.  This characteristic
does not apply in a peer-to-peer network where load-balancing updates
may have to travel across the Internet.

Most cluster algorithms assume that servers are homogeneous.  The
exceptions to this rule include work by Castro et al. \cite{castro99}.
This work assumes that servers will have different processing
capabilities and allows each server to stipulate a \emph{maximum
desirable utilization} that is incorporated into the load-balancing
algorithm.  The algorithm they use assumes that servers are
synchronized and send their load updates at the same time.  This is
not true in a peer-to-peer network where replicas cannot be
synchronized.  Zhu et al. \cite{zhu98} assume servers are
heterogeneous and use a metric that combines available disk capacity
and CUP cycles to choose a server within the cluster to handle a task
\cite{zhu98}.  Their algorithm uses a combination of random polling
before selection and random multicasting of load-balancing information
to a select few servers.  Both are techniques that would not scale in
a large peer-to-peer network.

Another well-studied load-balancing cluster approach is to have
heavily loaded servers handoff requests they receive to other servers
within the cluster that are less loaded or to have lightly loaded
servers attempt to get tasks from heavily loaded servers (e.g.,
\cite{dandamudi95,shivaratri90}). This can be achieved through
techniques such as HTTP redirection (e.g.,
\cite{cardellini99,andresen97,cardellini00}) or packet header
rewriting (e.g., \cite{aversa00}) or remote script execution
\cite{zhu98}.  HTTP redirection adds additional client round-trip
latency for every rescheduled request.  TCP/IP hand-off and packet
header rewriting require changes in the OS kernel or network interface
drivers.  Remote script execution requires trust between the serving
entities.

Similar to task handoff is the technique of process migration.
Process migration to spread job load across a set of servers in a
local-area distributed system has been widely studied both in the
theoretical literature as well as the systems literature (e.g.,
\cite{douglis91,luling93,downey95,petri95,lu96}). In these systems
overloaded servers migrate some of their currently running processes
to lighter loaded servers in an attempt to achieve more equitable
distribution of work across the servers.

Both task handoff and process migration require close coordination
amongst serving entities that can be afforded in a tightly-coupled
communication environment such as a cluster or local-area distributed
system.  In a peer-to-peer network where the replica nodes serving the
content may be widely distributed across the Internet, these
techniques are not possible.

A lot of work has looked at balancing load across multi-server
homogeneous web sites by leveraging the DNS service used to provide
the mapping between a web page's URL and the IP address of a web
server serving the URL.  Round-robin DNS was proposed, where the DNS
system maps requests to web servers in a round-robin fashion
\cite{katz94,andresen95}.  Because DNS mappings have a Time-to-Live
(TTL) field associated with them and tend to be cached at the local
name server in each domain, this approach can lead to a large number
of client requests from a particular domain getting mapped to the same
web server during the TTL period.  Thus, round-robin DNS achieves good
balance only so long as each domain has the same client request rate.
Moreover, round-robin load-balancing does not work in a heterogeneous
peer-to-peer context because each serving replica gets a uniform rate
of requests regardless of whether it can handle this rate.  Work that
takes into account domain request rate improves upon round-robin DNS and
is described by Colajanni et
al. \cite{colajanni97}.  

Colajanni et al. later extend this work to balance load across a set
of widely distributed heterogeneous web servers \cite{colajanni98}.
This work proposes the use of adaptive TTLs, where the TTL for a DNS
mapping is set inversely proportional to the domain's local client
request rate for the mapping of interest (as reported by the domain's
local name server).  The TTL is at the same time set to be
proportional to the chosen web server's maximum capacity.  So web
servers with high maximum capacity will have DNS mappings with longer
TTLs, and domains with low request rates will receive mappings with
longer TTLs.  Max-Cap, the algorithm proposed in this thesis, also
uses the maximum capacities of the serving replica nodes to allocate
requests proportionally.  The main difference is that in the work by
Colajanni et al., the root DNS scheduler acts as a centralized
dispatcher setting all DNS mappings and is assumed to know what the
request rate in the requesting domain is like.  In the peer-to-peer
case the authority node has no idea what the request rate throughout
the network is like, nor how large is the set of requesting nodes.

Lottery scheduling is another technique that, like Max-Cap, uses
proportional allocation.  This approach has been proposed in the
context of resource allocation within an operating system (the Mach
microkernel) \cite{waldspurger94}.  Client processes hold tickets that
give them access to particular resources in the operating
system. Clients are allocated resources by a centralized lottery
scheduler proportionally to the number of tickets they own and can
donate their tickets to other clients in exchange for tickets at a
later point.  Max-Cap is similar in that it allocates requests to a
replica node proportionally to the maximum capacity of the replica
node.  The main difference is that in Max-Cap the allocation decision
is completely distributed with no opportunity for exchange of
resources across replica nodes.

\section{Conclusions}
\label{Conclusions}

In this paper we examine the problem of load-balancing in a
peer-to-peer network where the goal is to distribute the demand for a
particular content fairly across the set of replica nodes that serve
that content.  Existing load-balancing algorithms proposed in the
distributed systems literature are not appropriate for a peer-to-peer
network.  We find that load-based algorithms do not handle the
heterogeneity that is typical in a peer-to-peer network.  We also find
that algorithms based on available capacity reports can suffer from
load oscillations even when the workload request rate is as low as
60\% of the total maximum capacities of replicas.

We propose and evaluate Max-Cap, a practical algorithm for
load-balancing.  Max-Cap handles heterogeneity, yet does not suffer
from oscillations when the workload rate is below 100\% of the total
maximum capacities of the replicas, adjusts better to very large
fluctuations in the workload and constantly changing replica sets, and
incurs less overhead than algorithms based on available capacity since
its reports are affected only by extraneous load on the replicas.  We
believe this makes Max-Cap a practical and elegant algorithm to apply
in peer-to-peer networks.

\section{Acknowledgments}
This research is supported by the Stanford Networking Reseach Center,
and by DARPA (contract N66001-00-C-8015).

The work presented here has benefited greatly from discussions with
Armando Fox and Rajeev Motwani.  We thank them for their invaluable
feedback.  We also thank Petros Maniatis for his detailed comments on
earlier drafts of this paper.

\small

\newcommand{\etalchar}[1]{$^{#1}$}

\begin{center}
    {\bf Appendix}
\end{center}

It should not surprise the reader that Inv-Load does not handle heterogeneity
since the same load at one replica may have a different effect on another
with a different maximum capacity.  However, surprisingly it turns out that
when replicas are homogeneous, the performance of Inv-Load and Max-Cap are comparable.

In this set of experiments, there are ten replicas, each of whose
maximum capacity we set at 10 requests per second for a total
maximum capacity of 100 requests per second. Queries are generated
according to a Poisson process with a lambda rate that is 80\%
the total maximum capacities of the replicas.

\begin{figure}
\centerline{\includegraphics[width=8cm]{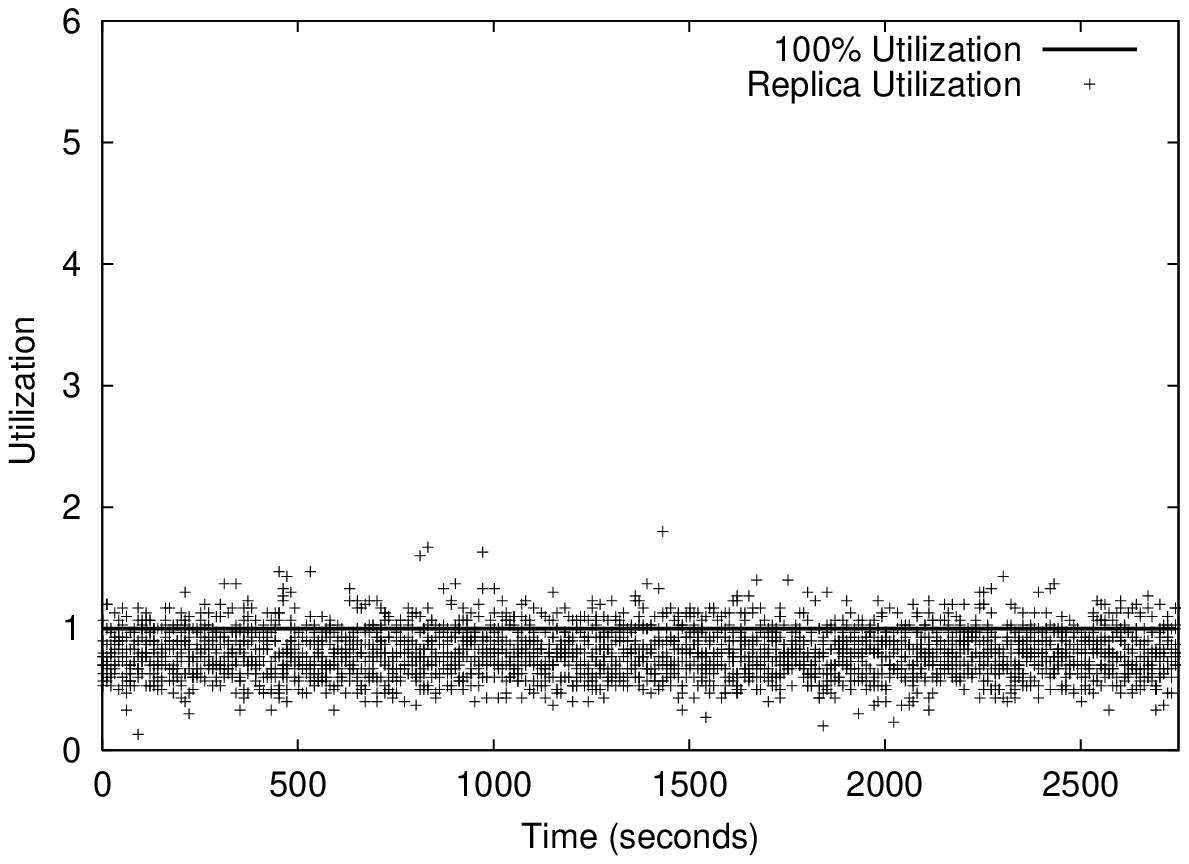}}
\caption{\small Replica Utilization versus Time for Inv-Load with an inter-update period of one second and homogeneous replicas.}
\label{LoadMaxEqCapsPoisson-80-a}
\end{figure}

\begin{figure}
\centerline{\includegraphics[width=8cm]{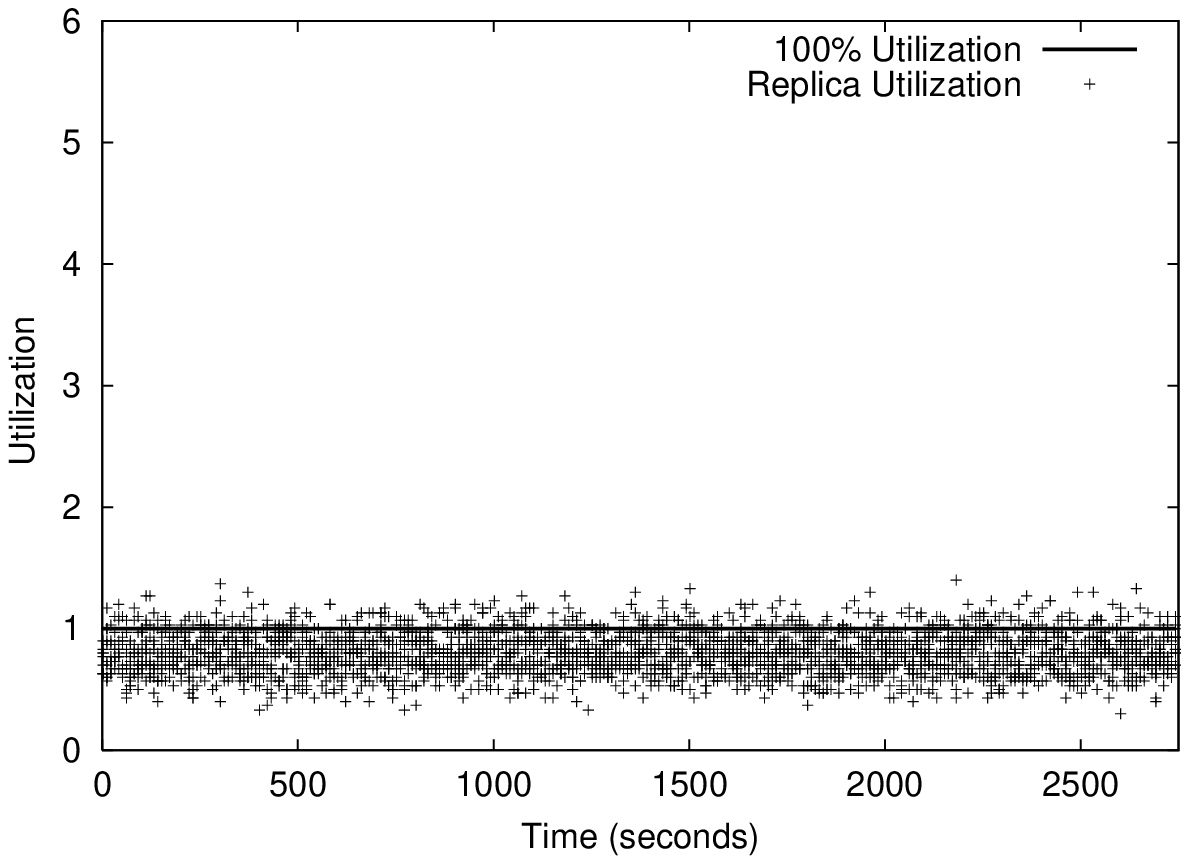}}
\caption{\small Replica Utilization versus Time for Max-Cap with homogeneous replicas.}
\label{LoadMaxEqCapsPoisson-80-b}
\end{figure}

Figures~\ref{LoadMaxEqCapsPoisson-80-a} and
\ref{LoadMaxEqCapsPoisson-80-b} show a scatterplot of how the
utilization of each replica proceeds with time when using Inv-Load
with a refresh period of one time unit and Max-Cap respectively.
Inv-Load and Max-Cap have similar scatterplots.

\begin{figure}
\centerline{\includegraphics[width=8cm]{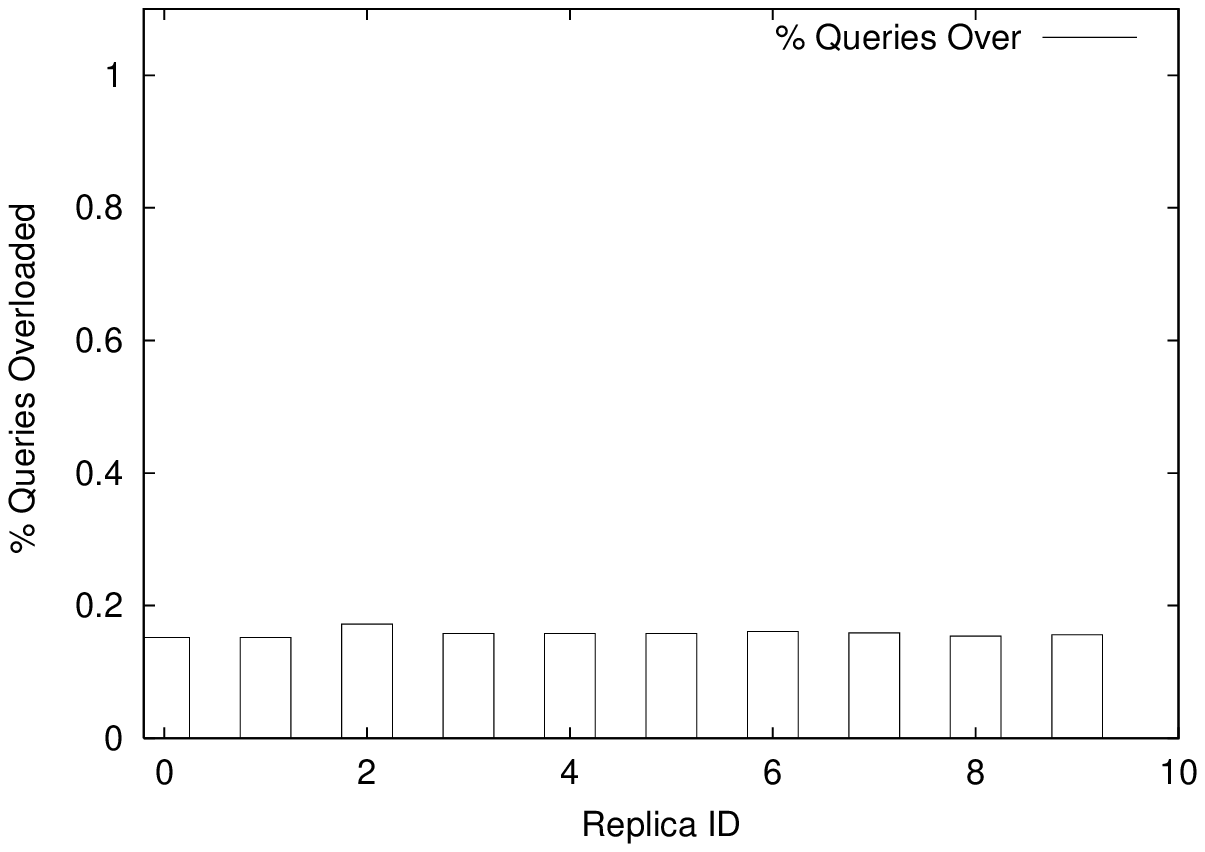}}
\caption{\small Percentage Overload Queries versus Replica ID for Inv-Load with an inter-update period of one second and homogeneous replicas.}
\label{LoadMaxEqCapsPoisson-80-OverCapQueries-a} 
\end{figure}

\begin{figure}
\centerline{\includegraphics[width=8cm]{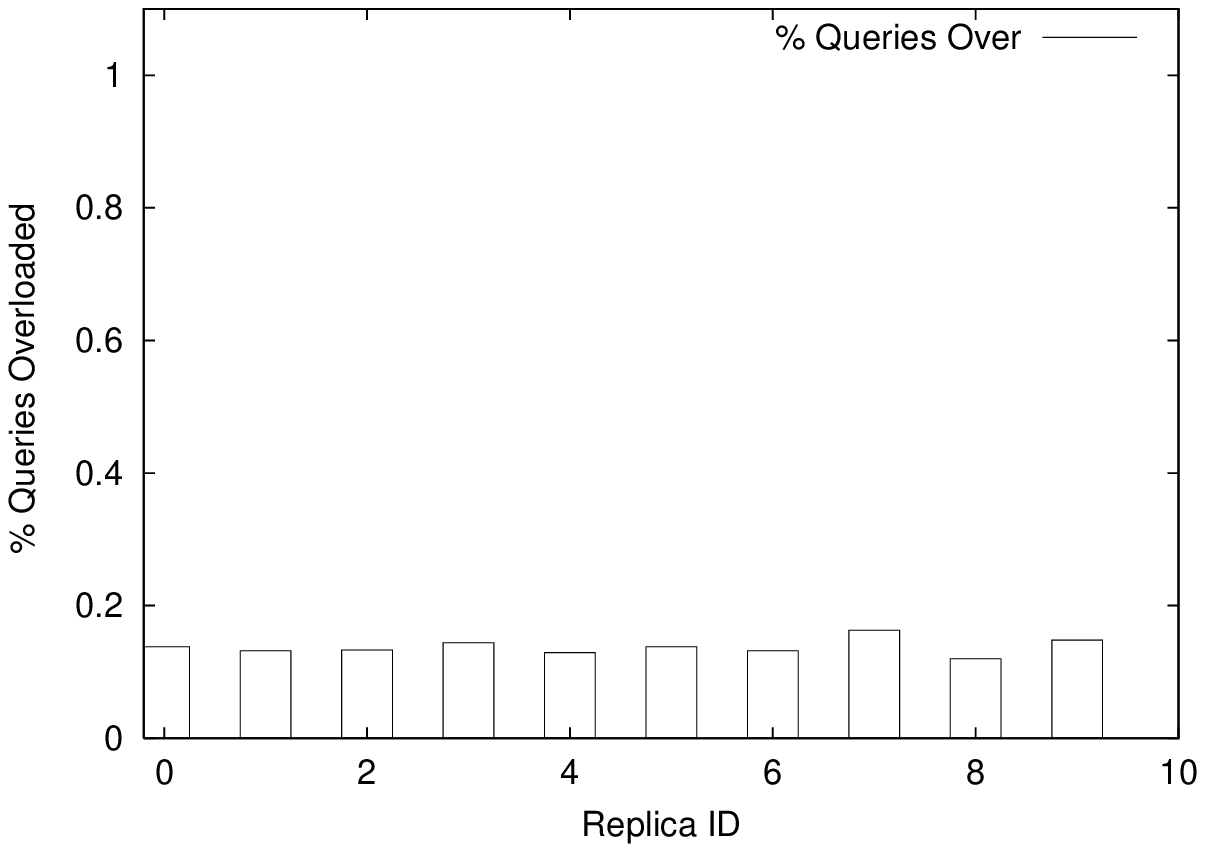}}
\caption{\small Percentage Overload Queries versus Replica ID for Max-Cap with homogeneous replicas.}
\label{LoadMaxEqCapsPoisson-80-OverCapQueries-b} 
\end{figure}

Figures~\ref{LoadMaxEqCapsPoisson-80-OverCapQueries-a} and
\ref{LoadMaxEqCapsPoisson-80-OverCapQueries-b} show for each replica,
the percentage of queries that arrived at the replica while the
replica was overloaded.  Again, we see that Inv-Load and Max-Cap have
comparable performance.

The difference is that Inv-Load incurs the extra overhead of one load
update per replica per second.  In a CUP tree of 100 nodes this
translates to 1000 updates per second being pushed down the CUP
tree. In a tree of 1000 nodes this translates to 10000 update per
second being pushed.  Thus, the larger the CUP tree, the larger the
overall network overhead.  The overhead incurred by Inv-Load could be
reduced by increasing the period between two consecutive updates at
each replica.  Increasing the period results in staler load updates.
We find that when experimenting with a range of periods (one to sixty
seconds), we confirm earlier studies \cite{mitzenmacher97} that have
found that as load information becomes more stale with increasing
periods, the performance of load-based balancing algorithms decreases.

We ran experiments with Pareto($\alpha$, $\kappa$) query interarrivals
with a wide range of $\alpha$ and $\kappa$ values (the Pareto
distribution shape and scale parameters) and found that with
homogeneous replicas, Inv-Load with a period of one and Max-Cap
continue to be comparable.  However, Max-Cap is preferable in these
cases because it incurs no overhead.

\end{document}